\RequirePackage{luatex85}
\documentclass[aps,pre,onecolumn,superscriptaddress,longbibliography,nofootinbib,showkeys,showpacs]{revtex4-1}
\usepackage{hyperref}
\hypersetup{
  pdfcreator = {},
  pdfproducer = {}
}

\usepackage[T1]{fontenc} 
\usepackage{mathpazo} 
\usepackage{amsfonts}
\usepackage{braket}
\usepackage{graphicx}
\graphicspath{ {./} }
\usepackage{mathptmx}      
\usepackage{bm}
\usepackage{enumerate}
\usepackage{subfigure}
\usepackage{color}
\usepackage{array}
\usepackage{tabularx}
\usepackage{lipsum}
\usepackage{tikz}
\usepackage{geometry}
\usepackage{qcircuit}
\usepackage{xcolor}
\usepackage{floatrow}
\usepackage{array}
\usepackage{tikzit}
\usepackage{pgfplots}
\usepgfplotslibrary{groupplots,dateplot}
\usetikzlibrary{patterns,shapes.arrows}
\pgfplotsset{compat=1.13}

\tikzstyle{new style 0}=[fill=red, draw=black, shape=circle]
\tikzstyle{color}=[fill=none, draw={rgb,255: red,49; green,49; blue,49}, shape=circle]
\tikzstyle{rectangle text}=[fill=none, draw=black, shape=rectangle]
\tikzstyle{new style 1}=[fill=none, draw=black, shape=circle]

\tikzstyle{light}=[-, fill=none, draw={rgb,255: red,189; green,189; blue,189}]
\tikzstyle{arrow}=[draw=black, ->]
\tikzstyle{doublearrow}=[<->]

\usepackage{tabularx}
\usepackage[autostyle=true]{csquotes} 
\usepackage{amsmath,amssymb}

\definecolor{DarkGreen}{RGB}{0,100,0}
\begin{document}
\title{Assessment of the variational quantum eigensolver:\\ 
application to the Heisenberg model}
	
\author{Manpreet Singh Jattana}
\affiliation{Institute for Advanced Simulation, J{\"u}lich Supercomputing Centre, Forschungszentrum J{\"u}lich, D-52425 J{\"u}lich, Germany}
\affiliation{RWTH Aachen University, D-52062 Aachen, Germany}

\author{Fengping Jin}
\affiliation{Institute for Advanced Simulation, J{\"u}lich Supercomputing Centre,\\ Forschungszentrum J{\"u}lich, D-52425 J{\"u}lich, Germany}

\author{Hans De Raedt}
\affiliation{Zernike Institute for Advanced Materials,\\
	University of Groningen,
	NL-9747 AG Groningen, The Netherlands}
\affiliation{Institute for Advanced Simulation, J{\"u}lich Supercomputing Centre,\\ Forschungszentrum J{\"u}lich, D-52425 J{\"u}lich, Germany}

\author{Kristel Michielsen}
\affiliation{Institute for Advanced Simulation, J{\"u}lich Supercomputing Centre,\\ Forschungszentrum J{\"u}lich, D-52425 J{\"u}lich, Germany}
\affiliation{RWTH Aachen University, D-52062 Aachen, Germany}
\date{\today}

\begin{abstract}
We present and analyze large-scale simulation results of a hybrid quantum-classical variational method to calculate the ground state energy of the anti-ferromagnetic Heisenberg model. Using a massively parallel universal quantum computer simulator, we observe that a low-depth-circuit ansatz advantageously exploits the efficiently preparable N\'{e}el initial state, avoids potential barren plateaus, and works for both one- and two-dimensional lattices. The analysis reflects the decisive ingredients required for a simulation by comparing different ans\"{a}tze, initial parameters, and gradient-based versus gradient-free optimizers. Extrapolation to the thermodynamic limit accurately yields the analytical value for the ground state energy, given by the Bethe ansatz. We predict that a fully functional quantum computer with 100 qubits can calculate the ground state energy with a relatively small error.
\end{abstract}

\maketitle
\section{Introduction}\label{INT}
Variational methods, in particular the variational quantum eigensolver (VQE) \cite{Peruzzo2014, McClean2016}, have recently been successful in demonstrating to solve proof-of-concept problems on current quantum computing devices \cite{OMalley2016, Kokail2019, Kandala2017}. Despite the initial success, it remains an open question which problems would demonstrate an advantage on future quantum computers. Finding the ground state energy of the Heisenberg model is one of the candidates.

Recent works have focused on the implementation of the VQE on quantum computers, including the invention of efficient methods for current devices \cite{Kandala2017, Huggins2019}, the reduction of the total number of required qubits \cite{Liu2019, Fujii2020}, the testing of optimization algorithms \cite{Koczor2020, Ostaszewski2021}, and a study of the effects of noise \cite{Zeng2020}. Also, attempts to implement the Bethe ansatz \cite{Bethe1931} on a quantum computer \cite{Nepomechie2020, VanDyke2021} have been made. Results for implementations of the VQE on quantum computers with up to 20 qubits \cite{Lyu2020} or less \cite{Seki2020, Slattery2021, Jin2020, Bespalova2020} are available.

Despite the progress in hybrid quantum-classical variational methods, several of their important aspects are still unexplored. Large-scale simulations of VQE for calculating the ground state energy of the Heisenberg model have not yet been performed. It is unclear if a single ansatz can be used for both the one- and two-dimensional lattices. A clear picture of how the minimum energy scales by using a certain ansatz within and beyond what is emulatable on classical hardware is missing. In this work, we present results for all these aspects.

The rest of the paper is structured as follows. In Sec.~\ref{sec2}, we briefly review the variational principle and the Heisenberg model, and introduce the ansatz. In Sec.~\ref{sec3}, we present the results of our work for one- and two-dimensional lattices. Finally, in Sec. \ref{sec4}, we summarise our findings.

\section{Theory}\label{sec2}
\subsection{Variational principle}
\begin{figure}
\includegraphics{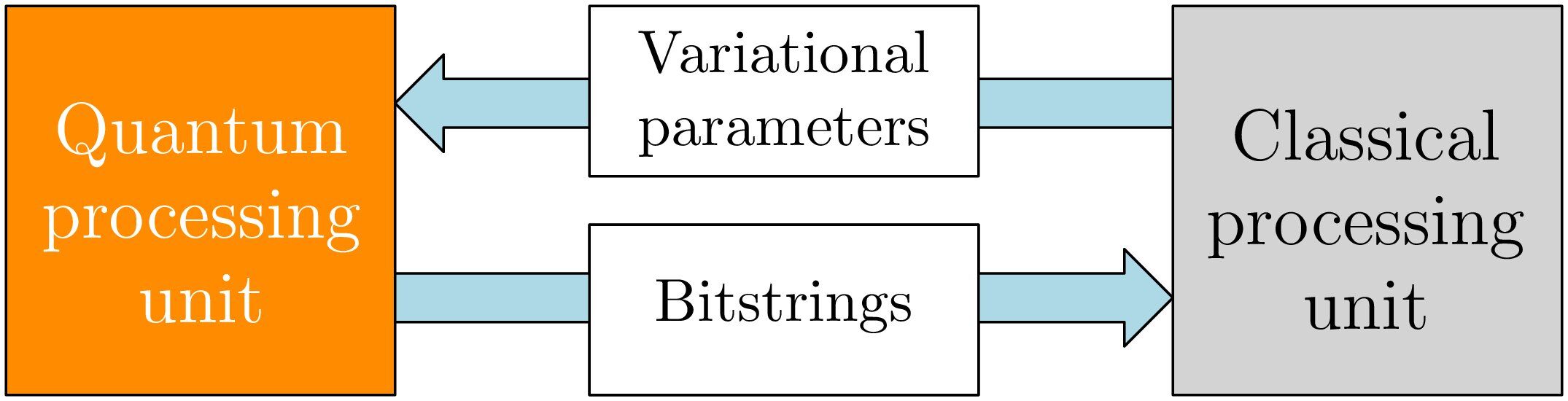}
\caption{(Colour online) Schematic of the hybrid quantum-classical mechanism of a variational quantum eigensolver.\label{fig11}}
\end{figure}

The variational principle states that the energy $E$ obtained by using a certain parameterized wavefunction $\psi(\theta)$ for a problem Hamiltonian $H$ is a strict upper bound to the ground state energy $E_0$ of $H$:
\begin{equation}
 E = \braket{\psi(\theta )|H|\psi(\theta ) } \geq E_0.\label{eq4:1}
\end{equation}
The VQE uses the variational principle, Eq.~(\ref{eq4:1}), to compute $E_0$ of $H$ on a quantum computer. The VQE algorithm is a hybrid quantum-classical algorithm that utilizes resources from quantum and classical computers in an iterative process. The diagram depicted in Fig.~\ref{fig11} shows the link between the quantum processing unit (QPU) and the classical processing unit (CPU). The QPU is responsible for carrying out the computation for a certain quantum circuit that generates the state $\psi(\theta)$, depending on a set of parameters, and returns the corresponding bitstrings to the CPU obtained after the measurement. The bitstrings are accumulated and processed by the classical unit, fed to an optimizer that suggests the next set of parameters that will lower the energy in successive iterations.

\subsection{Heisenberg model}
We analyze the Hamiltonian representing the quantum spin model
\begin{equation}
    H = \sum_{i>j}^N \Big({J}_{ij}^{xx} \sigma_i^x \cdot \sigma_j^x+{J}_{ij}^{yy} \sigma_i^y \cdot \sigma_j^y+{J}_{ij}^{zz} \sigma_i^z \cdot \sigma_j^z \Big ),\label{eq4:ham1}
\end{equation}
where $N$ denotes the number of spins, and $\sigma^x,\sigma^y,$ and $\sigma^z $ are the Pauli matrices. Throughout the rest of the paper, we use units such that $\hbar=1$ and $J$'s are dimensionless. If $J^{\alpha \alpha}_{ij} = 1$ for all $i,j=1,\dots,N$, and $\alpha\in   \{x,y,z\}$, we call $H$ the \textit{isotropic anti-ferromagnetic} Heisenberg model Hamiltonian.  If all coefficients ${J}_{ij}^{\alpha \alpha}$ are chosen randomly in the interval $(0,1]$, then we call $H$ the \textit{random} Hamiltonian. We use both open and periodic boundary conditions and map each spin to a qubit. For most of our simulations we consider bipartite spin lattices with a single exception of a (frustrated) triangular spin lattice. 

\subsection{Ansatz}
In general, the final state of a system acted upon by a parametrized ansatz can be written in the form
\begin{equation}
    \ket{\psi_f(\bm \theta)} = U(\bm \theta)\ket{\psi_0},\label{eqps}
\end{equation}
where $\bm \theta $ are the variational parameters, $U$ is the ansatz, and $\ket{\psi_0}$ denotes the initial state. In this paper we demonstrate that the following ansatz is sufficient to yield an accurate approximation to the ground state energy.
\begin{equation}
U(\bm \theta) = \Big [\prod_{ l =N-1 }^{1} \prod_{k = N}^{ l+1} U_{lk}(\theta_{lk}) \Big ] \Big [\prod_{ l =N-1 }^{1} \prod_{k = N}^{ l+1} U_{kl}(\theta_{kl}) \Big ],\label{eqxyxy1}
\end{equation}
where 
\begin{equation}
      U_{pq}(\theta_{pq}) =
    \begin{cases}
      e^{-i\theta_{pq} \sigma_p^y\sigma_q^x } & \text{if $p=N$ or $q=N$,}\\
      e^{-i\theta_{pq} \sigma_p^y\sigma_q^x\sigma_N^z } & \text{otherwise.}\label{eqxy1}
    \end{cases}   
\end{equation}
We elaborate on how to expand the Eq.~(\ref{eqxyxy1}). There is an independent index $l$ and a dependent index $k$. The index $l$ is monotonically decreased from $N-1$ to $1$. For each value of $l$, the dependent index $k$ is decreased from $N$ to $l+1$. The index $l$ is not decremented until all the values of $k$ are enumerated. These values of $l$ and $k$ describe the indices of the qubits that the operators in Eq.~(\ref{eqxy1}) act upon. In every operator, no more than three qubits are involved. The corresponding parameters $\theta_{kl}$ are independent for each unique combination of $l$ and $k$. The parameters $\theta_{kl}$ affect the phase of the $N^{\text{th}}$ qubit. The number of unitary operators in Eq.~(\ref{eqxyxy1}) is given by $N(N-1)$. The ordering of the unitary operators is important. In this paper, the specific combination of the operators in Eq.~(\ref{eqxy1}) is termed the XY-ansatz. Any other ordering is prone to produce results which may be different from each other. This is consistent with the findings in recent literature \cite{Grimsley2020, Tranter2019}.

The motivation to keep the number of operators to a maximum of two or three is inspired from the coupled cluster ansatz which can be powerful enough to express relevant states even in this restrictive form \cite{McClean2016}. It is then intuitive to try this approach also for the Heisenberg model. Accordingly, such an ansatz is expressed by
\begin{equation}
	\ket{\psi_f(\bm \theta)} = e^{-i A(\bm \theta)}\ket{\psi_0},\label{eq4:123}
\end{equation}
where $A$ can contain sums of products of Pauli operators. The implementation of an ansatz, e.g. the one given by Eq.~(\ref{eq4:123}), is not a simple task in general as it requires factorization of the matrix exponential $e^{-iA(\bm \theta)}$ \cite{DeRaedt1987}. Factorization creates a series of products of unitary operations, which results in deeper quantum circuits with a large number of gates. In this work, we do not directly implement the ansatz given by Eq.~(\ref{eq4:123}), but seek for other ans\"{a}tze in a factorized form which do not require further factorization. In effect, we create a quantum circuit from an set of operators instead of a sum of operators. Such an approach allows us to build low-depth quantum circuits. Furthermore, from an experimental perspective, it is difficult to build a quantum computing device in which all the qubits work equally well. Some qubits may perform certain gates more efficiently than others. In order to exploit such devices efficiently and to accommodate for experimental imperfections, we proposed the ansatz where all the parameterized gates are placed on only one qubit. All operations of the parametrised gates can be restricted to this single qubit.

For comparison with the XY-ansatz, we consider two different ans\"{a}tze. The first one is inspired by quantum chemistry. The unitary coupled cluster ansatz restricted to single and double excitations (UCCSD) is shown to produce results with chemical accuracy \cite{Xia2020, Romero2019, Barkoutsos2018, Shen2017}. We consider
\begin{equation}
    U(\bm \theta) = \prod_{l=N-1}^{1} \prod_{k=N}^{l+1}  U_{kl}(\theta_{kl}),\label{eqall}
\end{equation}
where
\begin{equation}
    U_{kl}(\theta_{kl}) = 
    \begin{cases}
      \prod_{\substack{\alpha}} \prod_{\beta} e^{-i  \theta_{kl}^{\beta \alpha} \sigma_k^\beta \sigma_l^\alpha } & \text{if $k=N$ or $l=N$,}\\
       \prod_{\alpha} \prod_{\beta} e^{-i \theta_{kl}^{\beta \alpha} \sigma_k^\beta \sigma_l^\alpha\sigma_N^z } & \text{otherwise,}
    \end{cases}   
\end{equation}
where $\alpha,\beta \in\{x,y,z \} $. A combinatorial calculation shows that the number of unitary operators in Eq.~(\ref{eqall}) is given by $3^2 N!/2!(N-2 )!$. Although the total number of terms scales polynomially rather than exponentially, further reductions are always welcome since the redundant terms often slow down the optimization process. Clearly, the operators in Eq.~(\ref{eqxyxy1}) are a subset of those in Eq.~(\ref{eqall}). The differences between using these two are highlighted in the results section.

The second ansatz is inspired by the problem Hamiltonian. We consider the ansatz which for the one-dimensional lattice Hamiltonians is given by 
\begin{equation}
    U(\bm \theta) = U_p(\bm \theta_p) \ldots U_1(\bm \theta_1), \label{1123}
\end{equation}
where $U_p$ is given 
\begin{equation}
    U_p (\bm \theta_p)= \prod_{k=1}^{N} U_{kp}(\bm \theta_{kp}), 
\end{equation}
and boundary conditions are used. Each $ U_{kp}(\bm \theta_{kp})$ is given by
\begin{equation}
    U_{kp} (\bm \theta_{kp})= 
        \begin{cases}
       \prod_{\alpha} e^{-i \theta_{k}^{\alpha} \sigma_k^{\alpha}\sigma_{k+1}^{\alpha} } & \text{if $k=N$ or $k+1=N$,}\\
      \prod_{\alpha} e^{-i \theta_{k}^{\alpha }\sigma_k^{\alpha}\sigma_{k+1}^{\alpha} \sigma_N^z } & \text{otherwise.}\label{eqh}
      \end{cases}
\end{equation}
For a lattice of size $N$ in one-dimension, the ansatz in Eq.~(\ref{1123}) has $p\times 3N$ unitary operators. The variational Hamiltonian ansatz \cite{Wecker2015} is itself inspired from adiabatic evolution. The idea is that a combination of ansatz and initial parameters that mimics the adiabatic evolution can have a lower initial energy to start the variational optimization. It has been used for solving the Hubbard-Fermi model \cite{Reiner2019, okokok} and a modified Haldane-Shastry Hamiltonian \cite{Wiersema2020}.

A good choice for the initial state $\ket{\psi_0}$ often yields better variational results. In the case of antiferromagnets, a good  $\ket{\psi_0}$ for the bipartite lattices is known to be the N\'{e}el state where one sublattice is initialised with spins anti-parallel to the other sublattice. The N\'{e}el state, for an even number of spins, is in the magnetic sector of zero magnetization, where the ground state for the one-dimensional isotropic anti-ferromagnetic Heisenberg model is located. For the frustrated lattice, half of the lattice spins are initialised anti-parallel to the other half, without regard to their location in the lattice. The qubits representing the spins for one group are initialised as zeros and those in the other group (anti-parallel) as ones.

\subsection{Implementation}

We use the massively parallel simulator J\"{u}lich Universal Quantum Computer Simulator (JUQCS) \cite{DeRaedt2007, DeRaedt2019} to  perform operations on the state vector. We also use Qiskit \cite{Qiskit} for small problem sizes. In an actual quantum device, the state vector itself is not accessible. Instead, the quantum device will produce an ensemble of bitstrings consisting of $0$s and $1$s only, from which the expectation values of the observables can be derived. This raises two issues. First, since the number of samples or bitstrings can only be finite, it is not always clear if finite sampling can accurately represent the underlying probability distribution. Second, we need a procedure to measure the individual terms of the Hamiltonian. While the first is an open problem, recent works have developed efficient methods for the second when it becomes a problem \cite{Bonet-Monroig2020, Verteletskyi2020, Hadfield2020, Gokhale2019, Huggins2021}. Fortunately, both these problems do not hinder finding the ground state energy of the Heisenberg model. First, on an actual quantum device, we do not need explicit knowledge of the probability distribution; one can calculate the expectation values directly by sampling. From the samples we can estimate the energy with an accuracy proportional to the square root of the number of samples. Second, unlike in quantum chemistry, the measurement of individual terms for the Heisenberg model is not a problem.

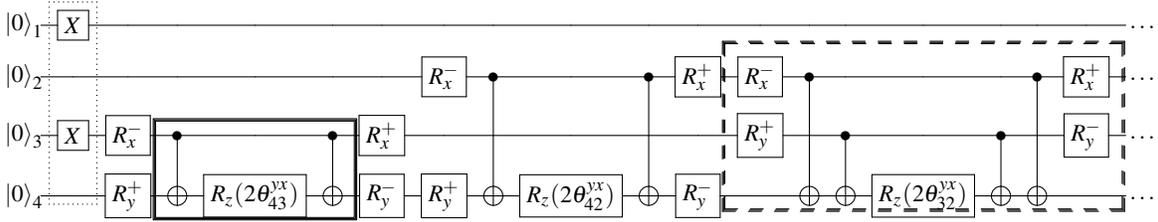
\begin{figure}
	\ffigbox
	{\caption{Circuit showing the first three parameters for a four-qubit XY-ansatz. The $R_x^+, R_x^-$, and $R_y^+, R_y^-$ gates are $R_x(\pi/2), R_x(-\pi/2)$, and $R_y(\pi/2), R_y(-\pi/2)$, respectively. The parameterized gate is always placed on the last qubit. Gates in the solid box correspond to the $e^{-i\theta \sigma^z \cdot \sigma^z}$ operator. Gates in the dashed box implement $e^{-i\theta \sigma^y \cdot \sigma^x \cdot \sigma^z}$.  The shown circuit is unoptimized and is optimized before usage. }\label{fig1}}
	{
		\[
		\Qcircuit @C=.7em @R=.7em {
		& \ket{ 0 }_1& &\gate{X}&\qw	& \qw & \qw          &  \qw	&\qw & \qw			&  \qw	&\qw&\qw	&\qw &\qw&\qw&\qw&\qw&\qw&\qw&\qw&\qw&\ldots\\
 & \ket{0  }_2&&\qw&\qw	& \qw      & \qw 		& \qw    &\qw 	& \gate{R_x^-} 			&	 \ctrl{2}		&\qw	&\ctrl{2} &\gate{R_x^+} &\gate{R_x^-}&\ctrl{2}&\qw&\qw&\qw&\ctrl{2} &\gate{R_x^+}&\qw&\ldots\\
&\ket{0}_3	&&\gate{X}&\gate{R_x^-}  & 	 \ctrl{1} 	&\qw 	& \ctrl{1} 	&  \gate{R_x^+}& \qw&\qw	&\qw&\qw	&\qw&\gate{R_y^+}&\qw&\ctrl{1} &\qw&\ctrl{1}&\qw&\gate{R_y^-}&\qw&\ldots\\
&\ket{0}_4	&&\qw&\gate{R_y^+} 	&  \targ   	    &\gate{R_z( 2\theta_{43}^{yx})}	& \targ  	& \gate{R_y^-}&	\gate{R_y^+}	& \targ	&\gate{R_z( 2\theta_{42}^{yx})}	&\targ& \gate{R_y^-}&\qw	&\targ& \targ&\gate{R_z( 2\theta_{32}^{yx})}&\targ& \targ&\qw \gategroup{1}{4}{4}{4}{.7em}{.} \gategroup{3}{6}{4}{8}{1.1em}{-}\gategroup{3}{6}{4}{8}{1.2em}{-}\gategroup{3}{6}{4}{8}{1.0em}{-} \gategroup{2}{15}{4}{21}{1.2em}{--}\gategroup{2}{15}{4}{21}{1.3em}{--}\gategroup{2}{15}{4}{21}{1.1em}{--}&\qw&\ldots\\
&\\
		}
		\]
	}
\end{figure}

The implementation of the ansatz is best understood through an example. Consider the four-qubit circuit shown in Fig.~\ref{fig1}, implementing the XY-ansatz for the $N=4$ isotropic Heisenberg ring (see Eq.~(\ref{eq4:ham1})). There are $12$ parameters in total, and Fig.~\ref{fig1} shows the implementation of the first three. After the initial state preparation, gates are applied to construct the unitary operators. The circuit in the rectangular solid box corresponds to the implementation of the $\exp(-i\theta_{43}^{yx} \sigma^z \cdot \sigma^z)$ operator having a variational parameter $ \theta_{43}^{yx}$. The terms containing $\sigma^x$ and $ \sigma^y$ are implemented by changing to the appropriate basis. The rectangular dashed box highlights the implementation of the third term in the ansatz, given by $\exp(-i \theta_{32}^{yx}  \sigma^y_3 \cdot \sigma^x_2 \cdot \sigma^z_4)$. Although the results for the special case of four qubits can be achieved using an even smaller subset of terms in the XY-ansatz operators containing only five parameters, for consistency and completeness, the twelve parameters are used for all the cases. The simulation for this example using Qiskit and JUQCS gives the energy $-8.0$, which is also the theoretical value obtained by exact diagonalization. The final state also has a $100\%$ overlap with the ground state. The corresponding z- and total-magnetization were both zero, as required.

For the classical optimisation part of the VQE, we use the sequential least squares quadratic programming (SLSQP) \cite{Kraft} optimizer in the SciPy package \cite{scipy}. SLSQP is a quasi-Newton gradient based algorithm. When using an ideal simulator, the gradients are computed numerically using the cost-effective forward differences formula. The calculation of the energy for given values of the parameters involves the quantum subroutine of VQE. Once the energies are obtained, the calculation of the gradient is done on the classical computer. Calculation of gradient and the next iterate is not a hard problem and classical computers can be used. It is the calculation of the energy for which quantum resources will be required. In our work, the cost of gradient computation for the optimizer is included in the total energy evaluations. For comparison, we also use a gradient-free optimizer called constrained optimization by linear approximation (COBYLA) \cite{scipy, Powell1994, Powell2007}.

\section{Results}\label{sec3}
\subsection{One-dimensional lattices}
\begin{figure}
\includegraphics[scale=0.96]{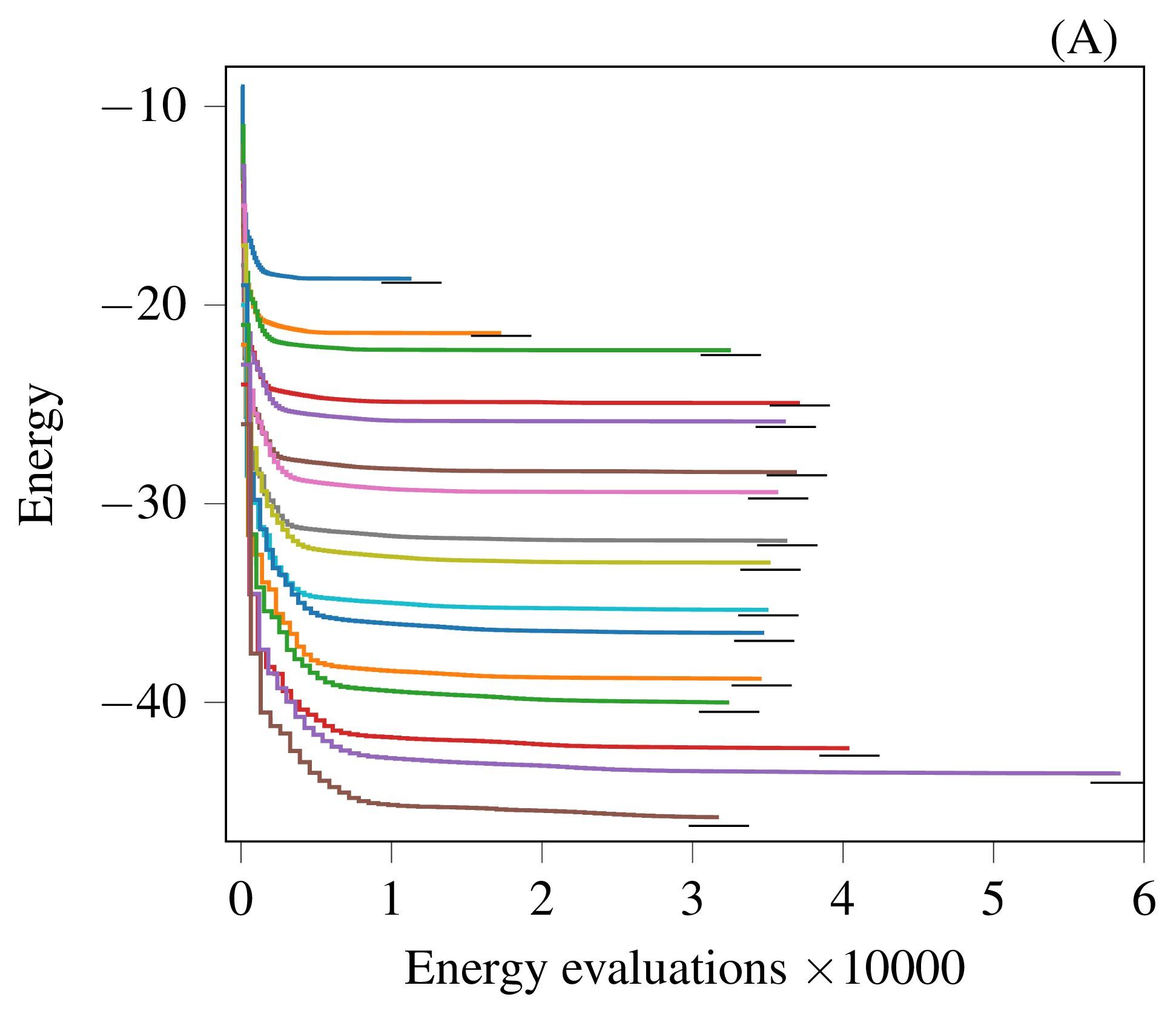}
\includegraphics[scale=0.96]{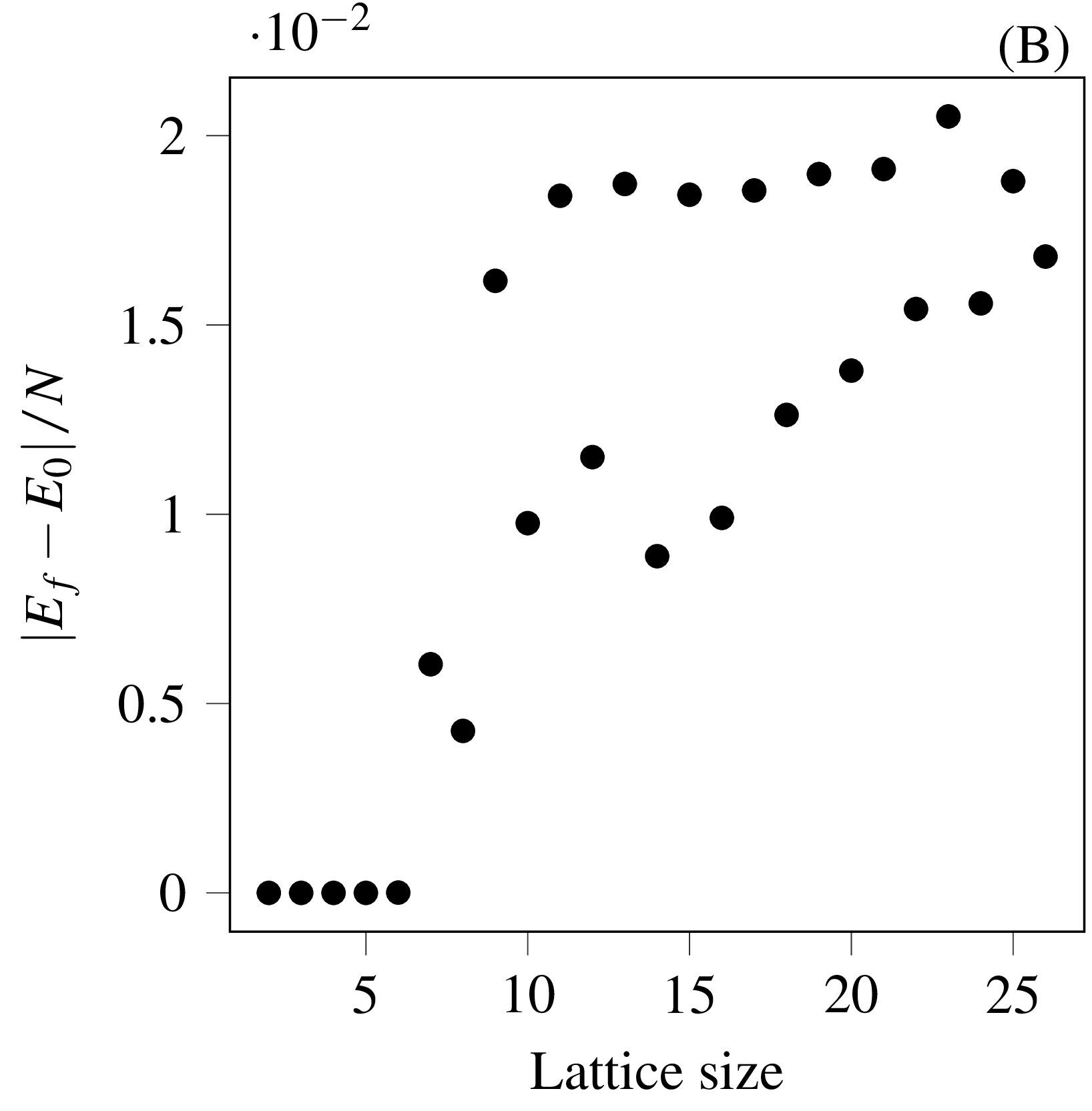}
	\caption{(Colour online) (A): Optimization progress using the XY ansatz for $11$ spins (first line), $12$ spins (second line) and so on, up to $26$ spins (bottom line). The small horizontal black lines represent the ground state energy. (B): Absolute difference in the variational and ground state energy per spin for different spins or lattice size. \label{fig2}}
\end{figure}

The results for the one-dimensional isotropic Hamiltonians with periodic boundary conditions are shown in Fig.~\ref{fig2}(A). The coloured lines show the optimization progress, i.e. the lowest energy achieved by successive energy evaluations for different lattice sizes. Initially, the system is in the N\'{e}el state. To take advantage of the N\'{e}el state, all the optimization parameters are initialised as zeros. As the optimization proceeds, the drop in energy is visible for all lattice sizes shown. The stepped progression is characteristic of the quasi-Newton optimization algorithms, which calculate the gradient before deciding on the step size. Each "step" of the staircase corresponds to a length equal to the number of parameters, since the formula of the forward differences for calculating the gradients requires $N(N-1)+1$ energy evaluations, where $N(N-1)$ is the number of parameters. Below each optimization curve is a horizontal black line corresponding to the energy of the ground state, obtained numerically by the Lanczos algorithm. According to the variational principle, the calculated energies are upper bounds to the energy of the ground states. Starting from a lattice with $14$ spins, the optimizer gave no signal for termination, but due to a time limit of $24$ hours on the supercomputer \cite{JUWELS}, the calculations stopped. Due to the different number of compute nodes, circuit depths, and parameters, the total number of energy evaluations differs for each lattice size. The energy optimization process can be restarted from the last values of the parameters. An example is shown for the lattice with $25$ spins, as can be seen from the longer curve resulting from the extra energy evaluations. In all the cases shown, the XY-ansatz produced a final energy $E_f$, which is close to the ground state energy. 

\begin{figure}
\includegraphics{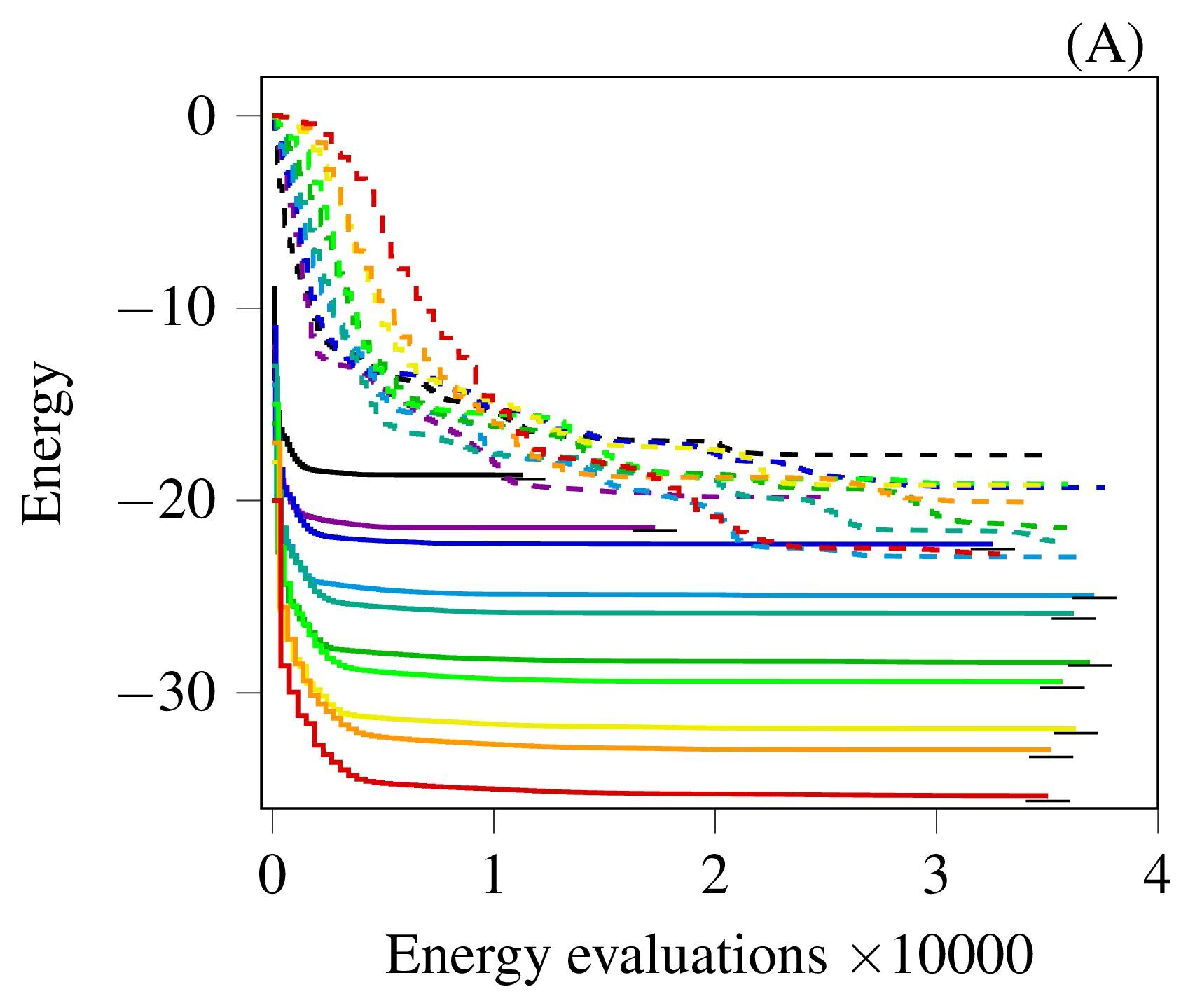}
\includegraphics{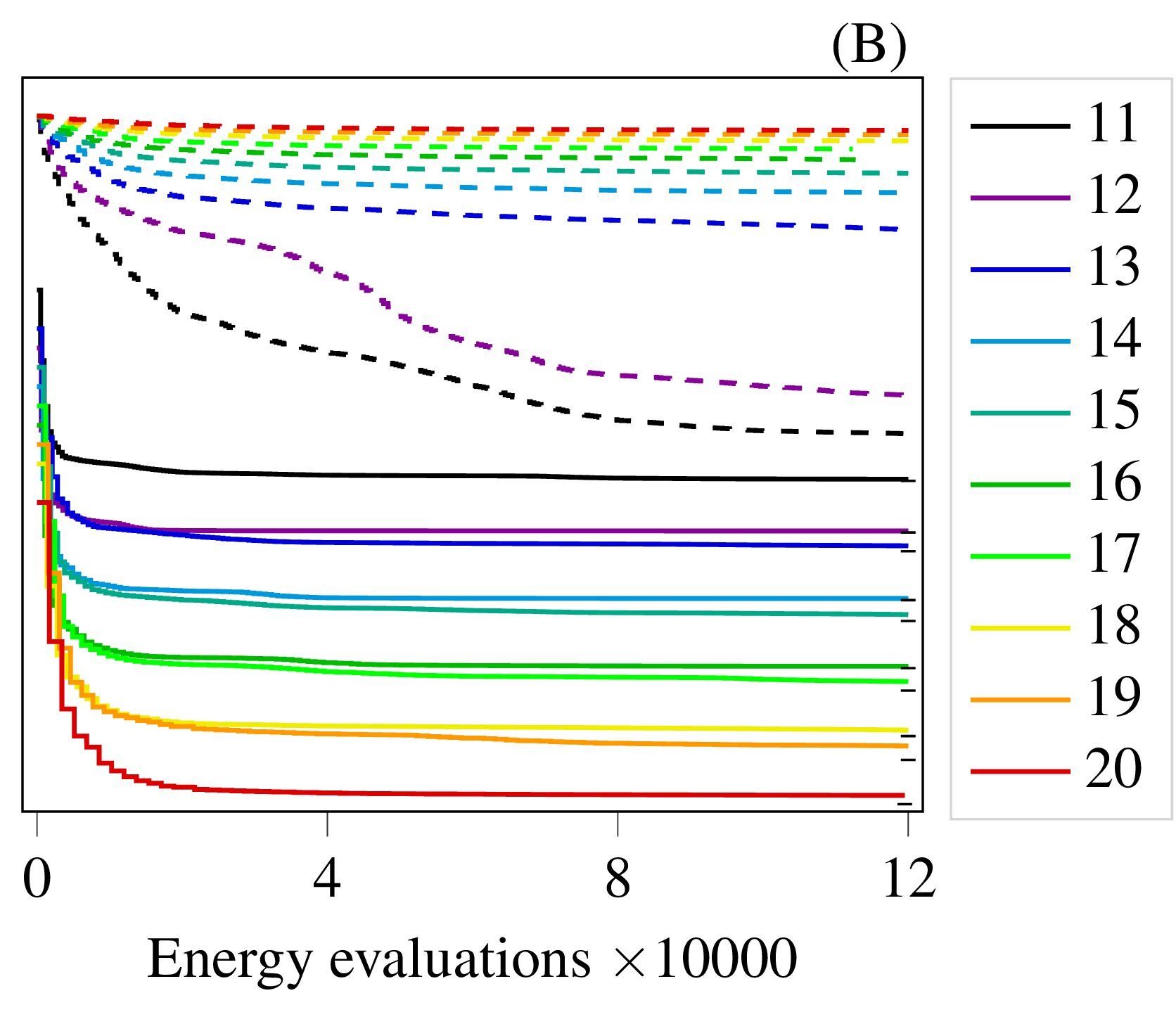}
	\caption{(Colour online) (A): Progress comparison when initializing with zeros (solid lines) versus random values (dashed lines) as parameters for the XY-ansatz. (B): Same as (A), except for using the ansatz given by Eq.~(\ref{eqall}). Legend depicts the lattice size for both (A) and (B). The small horizontal black lines represent the ground state energy. \label{figuccd} }
\end{figure}

The absolute difference between $E_f$ and the corresponding ground state energy $E_0$ per spin for each lattice size is shown in Fig.~\ref{fig2}(B). Results up to the lattice size of $6$ spins match exactly with the ground state energy (data not shown in (A)), and the differences show up for $N\geq 7$. The points cluster in two groups, one corresponding to lattices with an even and another one with an odd number of spins. We observe that the final energies for the lattices with an even number of spins are lower than for the ones with an odd number of spins. Additionally, we note that the ground states of the systems with an odd number of spins are degenerate. Through the variational principle, the ground state energies are mapped to the global minima of an energy landscape when an ansatz can express the ground state. There may be multiple global minima for degenerate cases. Due to this reason the energy landscape will be different from the even spin cases. Therefore, we conjecture that finding the ground state of degenerate cases using VQE is more challenging. This is also observed in the simulations we performed.

Figure~\ref{figuccd}(A) shows the energy optimization progress using two different sets of initial parameters. The legend indicates the lattice size. The lines can be separated into two groups, top (dashed) and bottom (solid). The bottom group corresponds to the case where the parameters were initialised as zeros, thus taking advantage of the N\'{e}el state and starting from a lower energy value. The top group of lines corresponds to parameters initialised as randomly distributed values in the interval $[0, 2\pi)$, not taking advantage of the N\'{e}el state and starting from a much higher energy, often close to $E = 0$. For all lattice sizes, initializing the parameters as zeros yields large drops in energy for the first few iterations, and $E_f$ is close to the ground state energy. After the significant drop, the (relative) progress slows down as the energy landscape becomes flatter near a minimum, and a large number of iterations is required to decrease the energy further. On the other hand, random initializations of the parameters do not yield significantly big drops in the energy in any of the cases. A prohibitively large number of energy evaluations seems required to obtain the same accuracy as when parameters are initialised as zeros. From a practical perspective, starting from random parameters does not appear to be very useful for the current problem.

Figure.~\ref{figuccd}(B) shows the energy optimization progress using the UCCSD inspired ansatz given by Eq.~(\ref{eqall}). Similar to plot (A), plot (B) shows the trend corresponding to the two different initializations of the parameters. Initializing all parameters as zeros is observed to have a significant initial drop in the energy, contrary to the random parameters. Interestingly, energy optimization progresses for cases with random initial parameters appear to be slower in proportion to the increasing lattice size, i.e. larger $N$ have a slower drop in energy. This effect appears to be consistent with what is termed in the recent literature as the \textit{barren-plateau} \cite{McClean2018,Campos2021,Holmes2021,Cerezo2020}. The larger lattice sizes, which require a larger number of parameters, lead to vanishingly small gradients. Given that vanishingly small gradients appear when one approaches a minimum, it is difficult to determine whether the random parameters landed in a local minimum of the energy landscape or at a barren-plateau. One way to find this out would be to restart multiple times with new sets of random parameters, but given that the barren-plateau effect is something that one aims to avoid, which is possible by initializing all parameters as zeros in this case, we skip such an approach.

In both plots (A) and (B) in Fig.~\ref{figuccd}, for parameters initialised as zeros, $E_f$ is very close to the ground state energy. This is understandable as the terms in the XY-ansatz are a subset of the terms given in Eq.~(\ref{eqall}). However, the difference between the two lies in the number of energy evaluations required to reach the ground state energy. The energies obtained with the XY-ansatz have the same (or comparable) values  as the energies obtained with the UCCSD inspired ansatz, but much less energy evaluations with fewer parameters.

\begin{figure}
\includegraphics{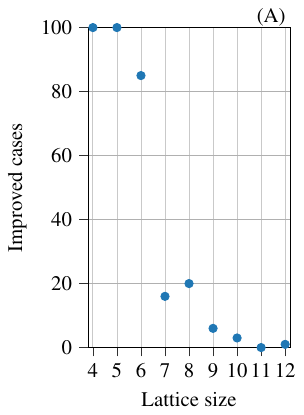}
\includegraphics{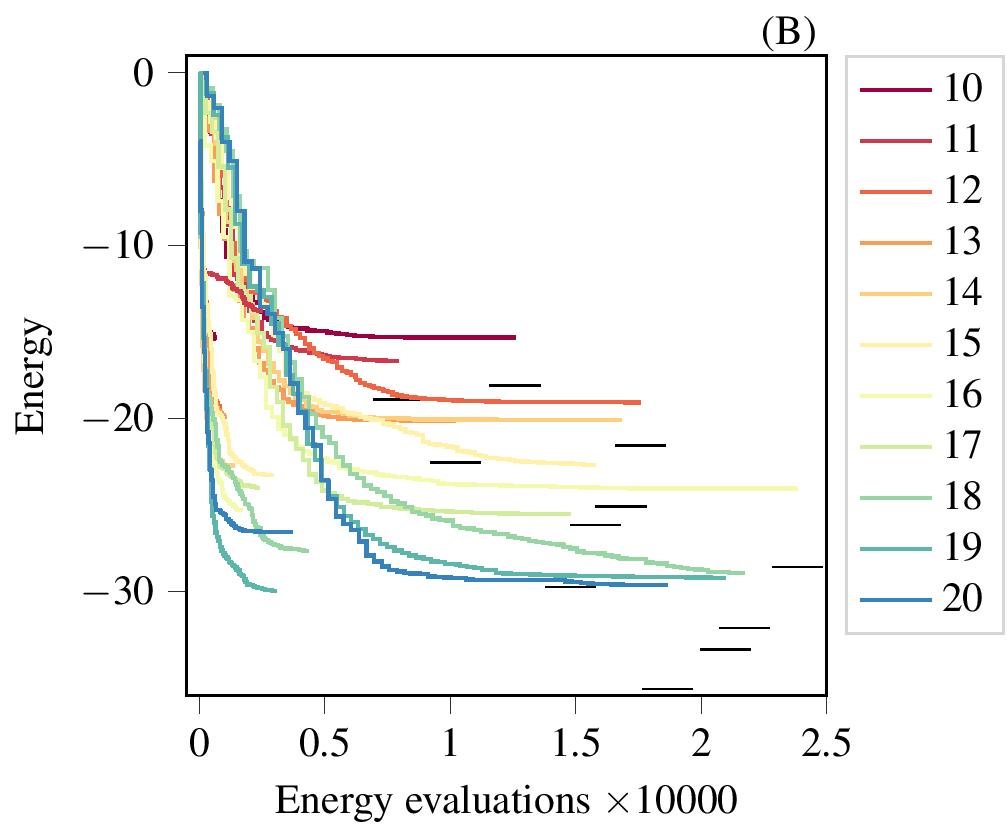}
	\caption{(Colour online) (A): The number of cases where the final energy obtained using random initial parameters was either equal to or lower than the energy obtained by setting zeros as initial parameters. (B): Optimization progress for the ansatz given by Eq. (9) for $p=1$ (group of shorter lines on the left) and $p=5$ (group of longer lines on the right). The legend shows the lattice size. The short horizontal black lines represent the ground state energy. \label{figh}}
\end{figure}

Despite the success of initializing the parameters as zeros, it should be noted that such an approach does not necessarily give the lowest possible energy given the ansatz. We performed random initializations with one hundred sets of random parameters for the smaller lattices. The results are shown in Fig.~\ref{figh}(A). We count and plot the number of cases in which the energy found by starting the optimiser using random initial parameters obtained a final energy equal to or lower than that obtained when starting from the N\'{e}el state. We observe that the cases drop sharpy as the lattice size increases. However, finding even a single energy lower than that found when starting from the N\'{e}el state shows that the minimum energy reached by initializing all parameters as zeros, although being very close to the ground state energy, is still only a local minimum and usually not a global minimum.

Figure~\ref{figh}(B) shows the energy optimization progress using the ansatz given by Eq.~(\ref{1123}). The lines can again be divided into two groups. The shorter lines on the left correspond to $p=1$ and the long lines on the right to $p=5$. The lines for $p=5$ are longer simply because in this case there are five times more parameters than for $p=1$, and so five times more energy evaluations are required to compute the gradient per iteration. For this ansatz, the parameters need to be initialised randomly since initializing them with zeros (for the tested cases with $N\leq 10$) results in the optimizer being trapped in a local minimum at the first iteration. It is observed that simply increasing the number of parameters in the circuit is no guarantee for improving the minimum energy. While for the lattice with $19$ spins a lower energy is reached with $p=1$ than with $p=5$, the opposite is the case for the lattice with $20$ spins. While a larger parameter space may facilitate a broader spectrum of states, the energy landscape may be the cause of this observation as local minima may surround the global minimum and impede reaching it. Since a random initialization would not be able to take advantage offered by the N\'{e}el state, and would thus require a much larger number of energy evaluations to converge to $E_f$, we restrict our study to only one such random initialization.

\begin{figure}
\includegraphics{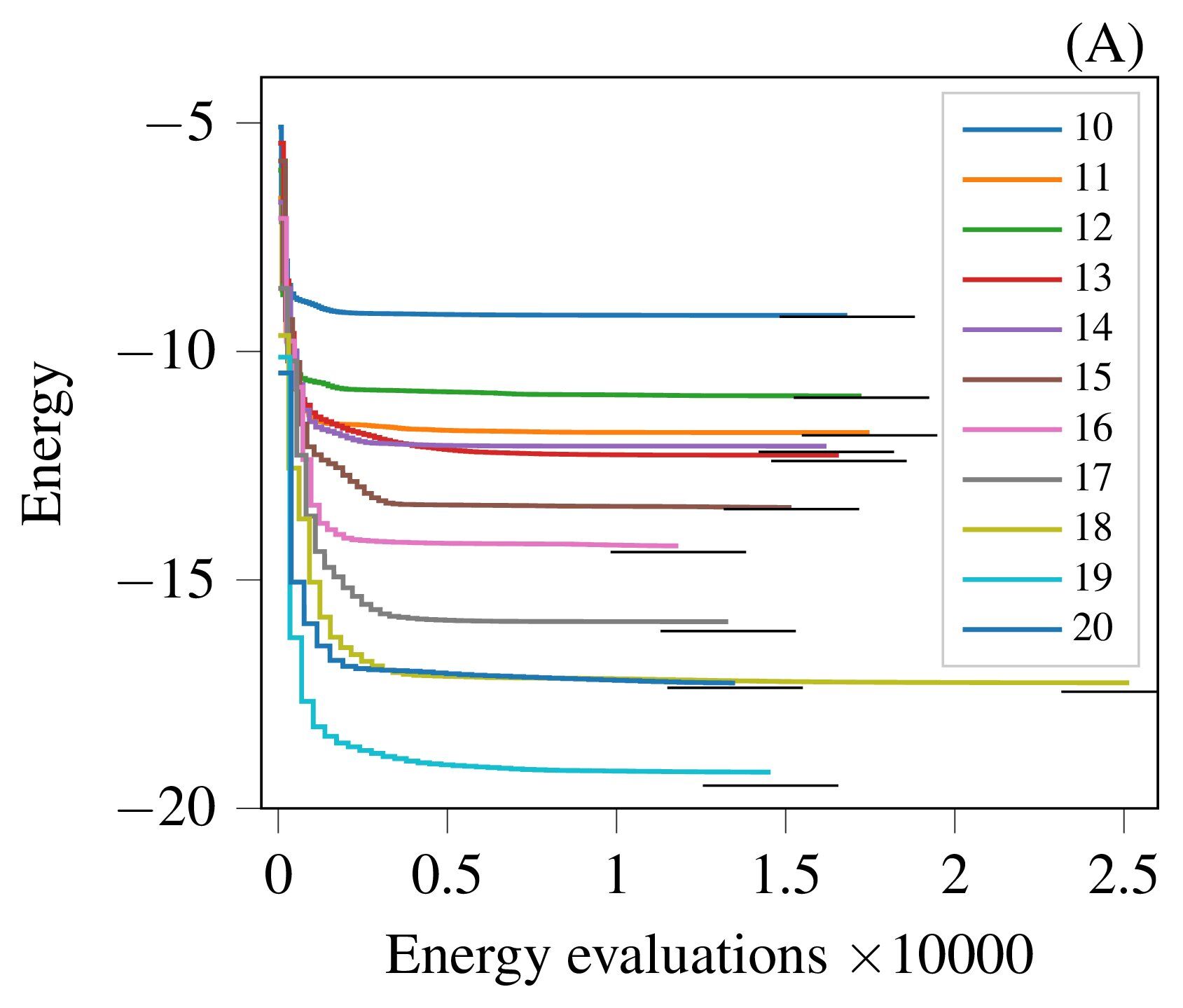}
\includegraphics{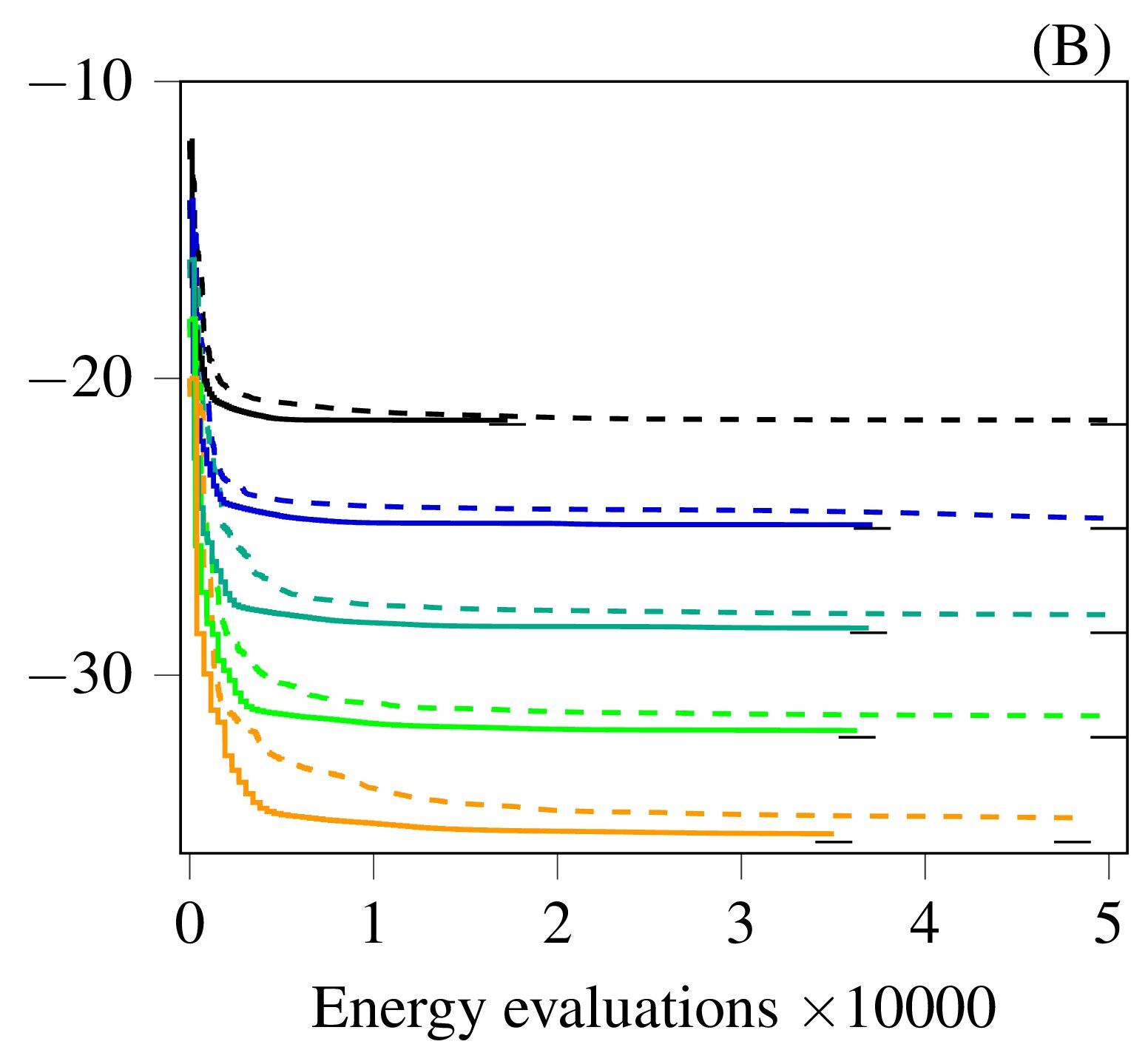}
	\caption{(Colour online) (A): Energy optimization progress using the XY-ansatz for different Hamiltonians with random coupling interactions for the different numbers of spins shown in legend. (B): Energy optimization progress using gradient-based (solid lines) and gradient-free (dashed lines) optimizers for rings  with random couplings between the spins with $12$ spins (first pair of the lines from top), $14$ spins (second pair of lines from the top), and so on up to $20$ spins (last pair of lines at the bottom). The short horizontal black lines represent the ground state energy.\label{figco}}
\end{figure}

Figure~\ref{figco}(A) shows the energy optimization progress for the random case of the anti-ferromagnetic one-dimensional ring. Using the XY-ansatz, the parameters were initialised as zeros. Except for the ring with $10$ spins, none of the optimization processes signalled convergence, and the optimization could be continued to reduce the energy further if required. A quick drop in the initial energy is also observed for the random case, and all the energies are reasonably close to the ground state energy.  

Figure~\ref{figco}(B) shows the energy optimization progress using the gradient-based optimizer SLSQP (solid lines) and the gradient-free optimizer COBYLA (dashed lines). We only show the results for the even lattice sizes. The results for the odd lattice sizes are similar. The gradient-based method gives lower energies at each energy evaluation and finds a much lower minimum faster than the gradient-free method. This result confirms the commonly accepted notion that for noiseless functions, gradient-based methods are superior.

\begin{figure}
	\includegraphics{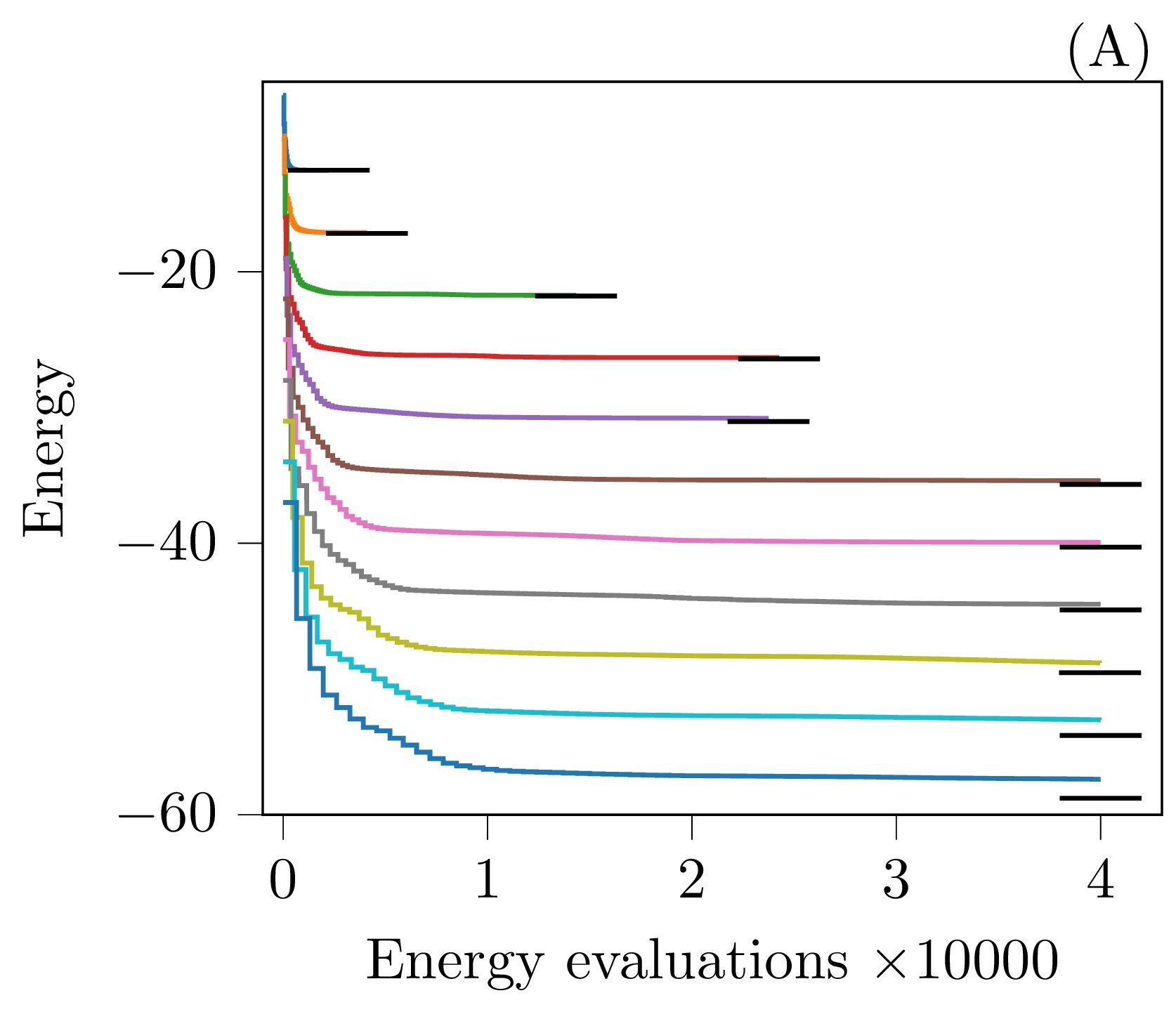}
	\includegraphics{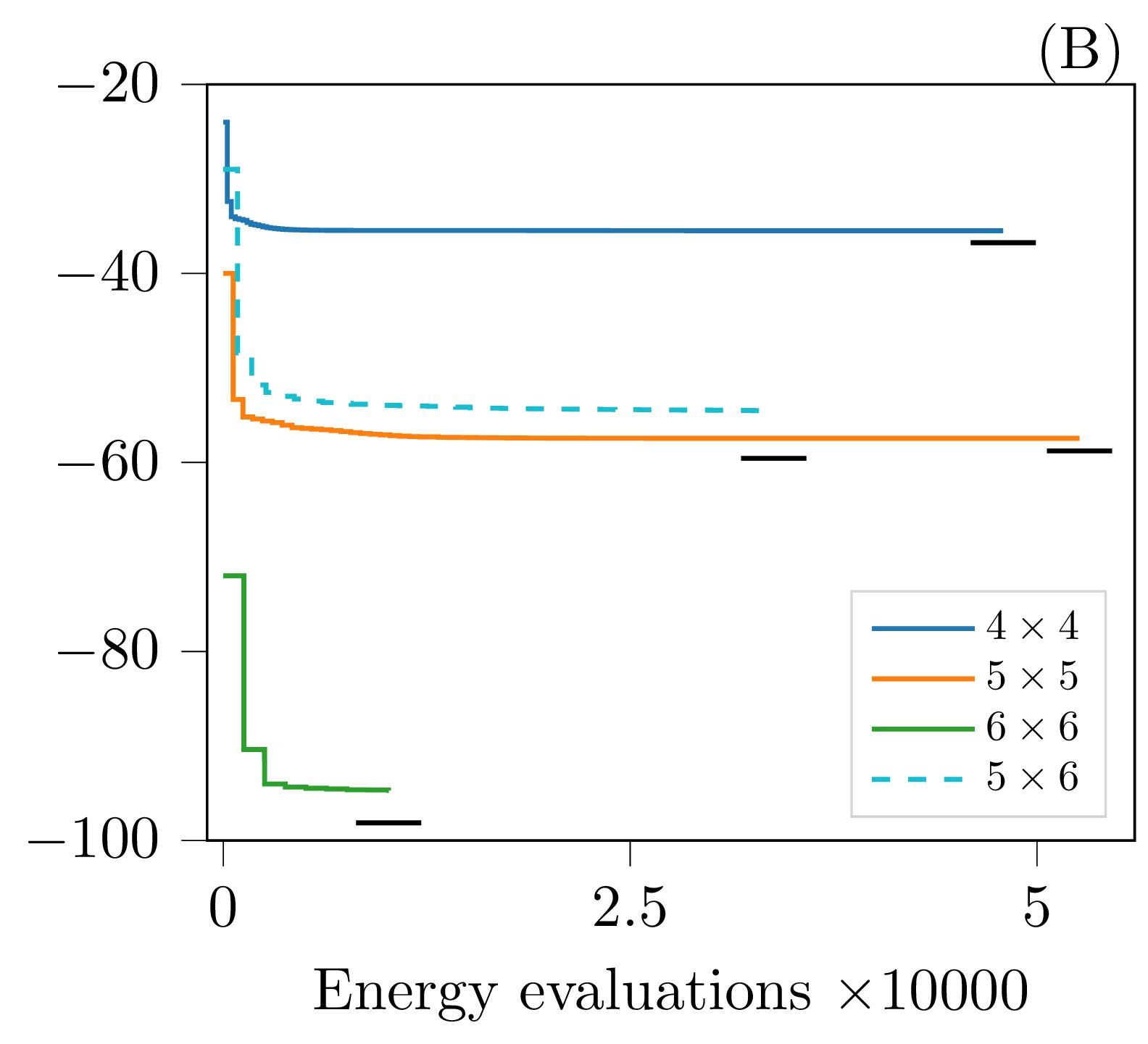}
	\caption{(Colour online) (A): Energy optimization progress for isotropic ladders with open boundary conditions. The top line is for a lattice with $3\times 2$ spins, the second for one with $4\times 2$ spins and so on until the bottom line which is for a lattice with $13\times 2$ spins. (B) Energy optimization progress for two-dimensional square lattices of size $4\times 4$ and $5\times 5$ with open boundary conditions and $6\times 6$ with periodic boundary conditions. The dashed line is for a frustrated triangular lattice of dimensions $5 \times 6$ with open boundary conditions.  The short horizontal black lines represent the ground state energy. \label{lad}}
\end{figure}

\subsection{Two-dimensional lattices}
We apply the XY-ansatz without any changes. Results for an isotropic ladder Hamiltonian with open boundary conditions are shown in Fig.~\ref{lad}(A). The results show the energy optimization progress for ladders of size $3\times 2$ up to size $13\times 2$. For example, the $8\times 2$ ladder, where the optimizer did not converge even after two runs of continued optimization, reported $E_f/N = -2.212$ as compared to the ground state energy per spin $E_0/N=-2.229$. 

Figure~\ref{lad}(B) shows results for three cases with square lattices and a $5\times 6$ (frustrated) triangular lattice.  The $4\times 4$ and $5\times 5$ lattices had open boundary conditions and the final energies were $E_f/N = -2.218$ and $E_f/N = -2.298$ as compared to $E_0/N=-2.297$ and $E_0/N=-2.351$, respectively. The optimized (unoptimized) circuit depths were $805$ $(1398)$ and $1939$ $(3531)$, respectively. For the frustrated lattice, $E_f/N = -1.817$ as compared to $E_0/N=-1.986$. The results for the frustrated lattice show a slightly larger gap in the energy obtained and the ground state. This could either be explained by assuming that the ansatz is less suitable for the case with the frustrated lattice or that the energy obtained when initializing all parameters as zeros corresponds to a local minimum far away from the global minimum.

The case of the square lattice of size $6\times 6$ with periodic boundary conditions poses a specific difficulty with the parameter optimization. The problem is that the XY-ansatz for $36$ spins has $1260$ parameters, and only one iteration can be performed as the evaluation of the gradient is possible only once within $24$ hours, the time per job on the supercomputer. Since the quasi-Newton methods require multiple iterations without losing the internal variables that systematically improve the convergence per iteration, the optimization is ill-suited for larger problems that cannot undergo multiple iterations in one run. However, this problem can be avoided by saving the internal variables of the optimizer, but this approach is beyond the scope of this work. Despite such a drawback, we proceeded with the first few iterations without saving the internal variables and were able to see a reasonable drop in the initial energy. The circuit depth after optimizing the circuit was reduced from $7458$ to $3985$. The final energy was $E_f/N = -2.631$ as compared to $E_0/N=-2.715$.

\begin{figure}
\includegraphics{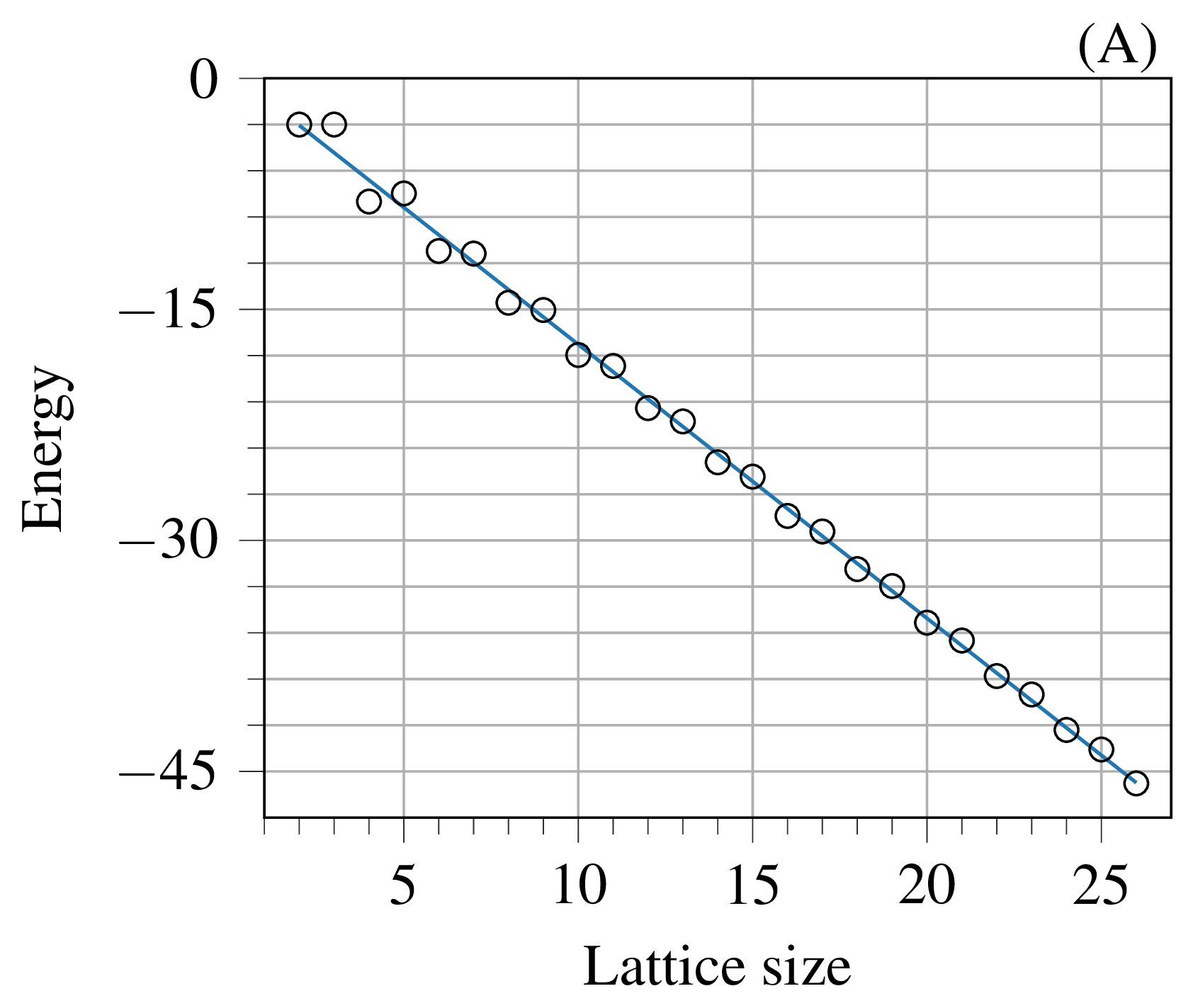}
\includegraphics{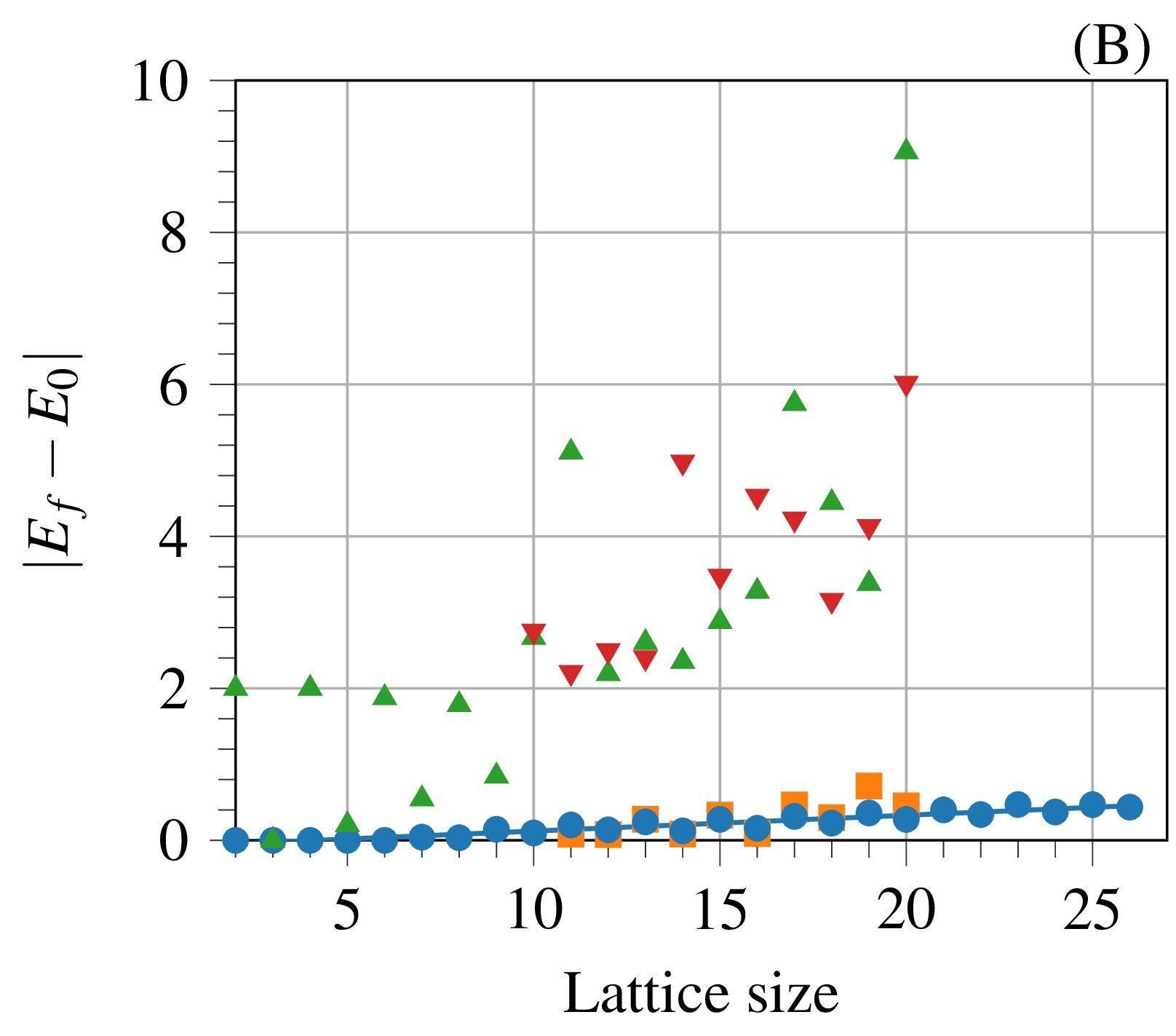}
	\caption{(Colour online) (A): Least squares fitting to the variational energies obtained for the one-dimensional isotropic rings of different sizes. The slope of the line, $-1.7783$, gives the energy for the infinite ring case. (B): Absolute differences of the variational and ground state energies for different lattice sizes for the isotropic ring using the ansatz from Eq.~(\ref{1123}) with $p=1$ (green triangles), $p=5$ (red inverted triangles), the ansatz from Eq.~(\ref{eqall}) (orange squares), and the XY-ansatz (blue dots). \label{figl}}
\end{figure}
\subsection{Extrapolation}
The energies obtained for the one-dimensional lattices can be fitted to a line for extrapolation, as shown in Fig.~\ref{figl}(A). The slope of the line that gives the energy per spin in the \textit{thermodynamic limit}, is $-1.7783$, which can be compared to the exact value $-1.7726$, known from the Bethe ansatz \cite{Parkinson2010}. The reported value only differs from the exact value by $5.7\times 10^{-3}$. Note that while the variational theorem guarantees that the energy obtained is a strict upper bound to the ground state, this is not necessarily the case when estimating the value in the thermodynamic limit by means of a fitted extrapolation. 

One measure of an ansatz's ability to scale up beyond what classical computers can simulate is to predict, given the available data, the expected difference between the VQE and exact energies. Such a calculation can be performed by extrapolating the available data. Figure.~\ref{figl}(B) shows data for four different ans\"{a}tze, namely the ansatz from Eq.~(\ref{1123}) with $p=1$ (green triangles), $p=5$ (red inverted triangles), the ansatz from Eq.~(\ref{eqall}) (orange squares), and the XY-ansatz (blue dots). Although some of the orange squares are lower than the corresponding blue dots, indicating that a lower energy was obtained  using the ansatz from Eq.~(\ref{eqall}) compared to using the XY-ansatz, the difference is small, and the XY-ansatz is the preferred choice because its number of parameters is significantly smaller. For both the $p$ values tested, the obtained data for the ansatz from Eq.~(\ref{1123}) puts it out of competition with the XY-ansatz. Moreover, there is no clear pattern that may help predict the behaviour for the ansatz from Eq.~(\ref{1123}) beyond the available data. Linear fitting is performed for the XY-ansatz, and the slope is equal to $2.0829 \times 10^{-2}$ with the intercept $-8.6202\times 10^{-2}$. Using the given slope, we predict that for a $100$-spin ring, the expected energy per spin will be higher than the ground state energy per spin by a value of approximately $0.02$. 

\section{Conclusions}\label{sec4}
We calculated the minimum energy for various implementations of the Heisenberg model for the one- and two-dimensional, isotropic, frustrated, and randomly-coupled lattices, using gradient-based and gradient-free optimizers, and different ans\"{a}tze. The herein proposed XY-ansatz shows reasonable results if all the simulation variables are optimized, i.e. a suitable initial state combined with a good quality optimizer and a good choice of initial parameters. 

Given an ansatz, there is currently no analytical method to ascertain if its global minimum corresponds to the ground state energy. Thus, we rely on optimization algorithms to navigate through the multi-dimensional rugged energy landscapes. In many such landscapes, there appear multiple local minima close to the ground state energy. For the cases where the exact ground state energy was not obtained using the XY-ansatz, it remains an open question if the global minimum of the energy landscape corresponds to the ground state energy. Thus, an improvement of the optimization algorithms appears to be essential for the further success of hybrid variational methods. 

For the variational simulations performed in this work, initializing the variables as zeros instead of random numbers produced better results. For the anti-ferromagnetic Heisenberg model, it is known that the N\'{e}el state is an efficient initial state, and starting from zeros takes advantage of this knowledge. Therefore, in general, it is the knowledge or insight about a particular problem Hamiltonian that is relevant for an improved performance of the variational methods. 

\section{Acknowledgement}	
The authors gratefully acknowledge the Gauss Centre for Supercomputing e.V. (www.gauss-centre.eu) for funding this project by providing computing time \cite{JUWELS}.  M.S.J.  acknowledges support from the project OpenSuperQ (820363) of the EU Quantum Flagship.
\bibliographystyle{apsrev4-1}  
\bibliography{heisenberg_model.bib}  

\begin{thebibliography}{48}%
\makeatletter
\providecommand \@ifxundefined [1]{%
 \@ifx{#1\undefined}
}%
\providecommand \@ifnum [1]{%
 \ifnum #1\expandafter \@firstoftwo
 \else \expandafter \@secondoftwo
 \fi
}%
\providecommand \@ifx [1]{%
 \ifx #1\expandafter \@firstoftwo
 \else \expandafter \@secondoftwo
 \fi
}%
\providecommand \natexlab [1]{#1}%
\providecommand \enquote  [1]{``#1''}%
\providecommand \bibnamefont  [1]{#1}%
\providecommand \bibfnamefont [1]{#1}%
\providecommand \citenamefont [1]{#1}%
\providecommand \href@noop [0]{\@secondoftwo}%
\providecommand \href [0]{\begingroup \@sanitize@url \@href}%
\providecommand \@href[1]{\@@startlink{#1}\@@href}%
\providecommand \@@href[1]{\endgroup#1\@@endlink}%
\providecommand \@sanitize@url [0]{\catcode `\\12\catcode `\$12\catcode
  `\&12\catcode `\#12\catcode `\^12\catcode `\_12\catcode `\%12\relax}%
\providecommand \@@startlink[1]{}%
\providecommand \@@endlink[0]{}%
\providecommand \url  [0]{\begingroup\@sanitize@url \@url }%
\providecommand \@url [1]{\endgroup\@href {#1}{\urlprefix }}%
\providecommand \urlprefix  [0]{URL }%
\providecommand \Eprint [0]{\href }%
\providecommand \doibase [0]{http://dx.doi.org/}%
\providecommand \selectlanguage [0]{\@gobble}%
\providecommand \bibinfo  [0]{\@secondoftwo}%
\providecommand \bibfield  [0]{\@secondoftwo}%
\providecommand \translation [1]{[#1]}%
\providecommand \BibitemOpen [0]{}%
\providecommand \bibitemStop [0]{}%
\providecommand \bibitemNoStop [0]{.\EOS\space}%
\providecommand \EOS [0]{\spacefactor3000\relax}%
\providecommand \BibitemShut  [1]{\csname bibitem#1\endcsname}%
\let\auto@bib@innerbib\@empty
\bibitem [{\citenamefont {Peruzzo}\ \emph {et~al.}(2014)\citenamefont
  {Peruzzo}, \citenamefont {McClean}, \citenamefont {Shadbolt}, \citenamefont
  {Yung}, \citenamefont {Zhou}, \citenamefont {Love}, \citenamefont
  {Aspuru-Guzik},\ and\ \citenamefont {O'Brien}}]{Peruzzo2014}%
  \BibitemOpen
  \bibfield  {author} {\bibinfo {author} {\bibfnamefont {A.}~\bibnamefont
  {Peruzzo}}, \bibinfo {author} {\bibfnamefont {J.}~\bibnamefont {McClean}},
  \bibinfo {author} {\bibfnamefont {P.}~\bibnamefont {Shadbolt}}, \bibinfo
  {author} {\bibfnamefont {M.~H.}\ \bibnamefont {Yung}}, \bibinfo {author}
  {\bibfnamefont {X.~Q.}\ \bibnamefont {Zhou}}, \bibinfo {author}
  {\bibfnamefont {P.~J.}\ \bibnamefont {Love}}, \bibinfo {author}
  {\bibfnamefont {A.}~\bibnamefont {Aspuru-Guzik}}, \ and\ \bibinfo {author}
  {\bibfnamefont {J.~L.}\ \bibnamefont {O'Brien}},\ }\href {\doibase
  10.1038/ncomms5213} {\bibfield  {journal} {\bibinfo  {journal} {Nature
  Communications}\ }\textbf {\bibinfo {volume} {5}},\ \bibinfo {pages} {1}
  (\bibinfo {year} {2014})}\BibitemShut {NoStop}%
\bibitem [{\citenamefont {McClean}\ \emph {et~al.}(2016)\citenamefont
  {McClean}, \citenamefont {Romero}, \citenamefont {Babbush},\ and\
  \citenamefont {Aspuru-Guzik}}]{McClean2016}%
  \BibitemOpen
  \bibfield  {author} {\bibinfo {author} {\bibfnamefont {J.~R.}\ \bibnamefont
  {McClean}}, \bibinfo {author} {\bibfnamefont {J.}~\bibnamefont {Romero}},
  \bibinfo {author} {\bibfnamefont {R.}~\bibnamefont {Babbush}}, \ and\
  \bibinfo {author} {\bibfnamefont {A.}~\bibnamefont {Aspuru-Guzik}},\ }\href
  {\doibase 10.1088/1367-2630/18/2/023023} {\bibfield  {journal} {\bibinfo
  {journal} {New Journal of Physics}\ }\textbf {\bibinfo {volume} {18}},\
  \bibinfo {pages} {023023} (\bibinfo {year} {2016})}\BibitemShut {NoStop}%
\bibitem [{\citenamefont {O'Malley}\ \emph {et~al.}(2016)\citenamefont
  {O'Malley}, \citenamefont {Babbush}, \citenamefont {Kivlichan}, \citenamefont
  {Romero}, \citenamefont {McClean}, \citenamefont {Barends}, \citenamefont
  {Kelly}, \citenamefont {Roushan}, \citenamefont {Tranter}, \citenamefont
  {Ding}, \citenamefont {Campbell}, \citenamefont {Chen}, \citenamefont {Chen},
  \citenamefont {Chiaro}, \citenamefont {Dunsworth}, \citenamefont {Fowler},
  \citenamefont {Jeffrey}, \citenamefont {Lucero}, \citenamefont {Megrant},
  \citenamefont {Mutus}, \citenamefont {Neeley}, \citenamefont {Neill},
  \citenamefont {Quintana}, \citenamefont {Sank}, \citenamefont {Vainsencher},
  \citenamefont {Wenner}, \citenamefont {White}, \citenamefont {Coveney},
  \citenamefont {Love}, \citenamefont {Neven}, \citenamefont {Aspuru-Guzik},\
  and\ \citenamefont {Martinis}}]{OMalley2016}%
  \BibitemOpen
  \bibfield  {author} {\bibinfo {author} {\bibfnamefont {P.~J.~J.}\
  \bibnamefont {O'Malley}}, \bibinfo {author} {\bibfnamefont {R.}~\bibnamefont
  {Babbush}}, \bibinfo {author} {\bibfnamefont {I.~D.}\ \bibnamefont
  {Kivlichan}}, \bibinfo {author} {\bibfnamefont {J.}~\bibnamefont {Romero}},
  \bibinfo {author} {\bibfnamefont {J.~R.}\ \bibnamefont {McClean}}, \bibinfo
  {author} {\bibfnamefont {R.}~\bibnamefont {Barends}}, \bibinfo {author}
  {\bibfnamefont {J.}~\bibnamefont {Kelly}}, \bibinfo {author} {\bibfnamefont
  {P.}~\bibnamefont {Roushan}}, \bibinfo {author} {\bibfnamefont
  {A.}~\bibnamefont {Tranter}}, \bibinfo {author} {\bibfnamefont
  {N.}~\bibnamefont {Ding}}, \bibinfo {author} {\bibfnamefont {B.}~\bibnamefont
  {Campbell}}, \bibinfo {author} {\bibfnamefont {Y.}~\bibnamefont {Chen}},
  \bibinfo {author} {\bibfnamefont {Z.}~\bibnamefont {Chen}}, \bibinfo {author}
  {\bibfnamefont {B.}~\bibnamefont {Chiaro}}, \bibinfo {author} {\bibfnamefont
  {A.}~\bibnamefont {Dunsworth}}, \bibinfo {author} {\bibfnamefont {A.~G.}\
  \bibnamefont {Fowler}}, \bibinfo {author} {\bibfnamefont {E.}~\bibnamefont
  {Jeffrey}}, \bibinfo {author} {\bibfnamefont {E.}~\bibnamefont {Lucero}},
  \bibinfo {author} {\bibfnamefont {A.}~\bibnamefont {Megrant}}, \bibinfo
  {author} {\bibfnamefont {J.~Y.}\ \bibnamefont {Mutus}}, \bibinfo {author}
  {\bibfnamefont {M.}~\bibnamefont {Neeley}}, \bibinfo {author} {\bibfnamefont
  {C.}~\bibnamefont {Neill}}, \bibinfo {author} {\bibfnamefont
  {C.}~\bibnamefont {Quintana}}, \bibinfo {author} {\bibfnamefont
  {D.}~\bibnamefont {Sank}}, \bibinfo {author} {\bibfnamefont {A.}~\bibnamefont
  {Vainsencher}}, \bibinfo {author} {\bibfnamefont {J.}~\bibnamefont {Wenner}},
  \bibinfo {author} {\bibfnamefont {T.~C.}\ \bibnamefont {White}}, \bibinfo
  {author} {\bibfnamefont {P.~V.}\ \bibnamefont {Coveney}}, \bibinfo {author}
  {\bibfnamefont {P.~J.}\ \bibnamefont {Love}}, \bibinfo {author}
  {\bibfnamefont {H.}~\bibnamefont {Neven}}, \bibinfo {author} {\bibfnamefont
  {A.}~\bibnamefont {Aspuru-Guzik}}, \ and\ \bibinfo {author} {\bibfnamefont
  {J.~M.}\ \bibnamefont {Martinis}},\ }\href {\doibase
  10.1103/PhysRevX.6.031007} {\bibfield  {journal} {\bibinfo  {journal} {Phys.
  Rev. X}\ }\textbf {\bibinfo {volume} {6}},\ \bibinfo {pages} {031007}
  (\bibinfo {year} {2016})}\BibitemShut {NoStop}%
\bibitem [{\citenamefont {Kokail}\ \emph {et~al.}(2019)\citenamefont {Kokail},
  \citenamefont {Maier}, \citenamefont {van Bijnen}, \citenamefont {Brydges},
  \citenamefont {Joshi}, \citenamefont {Jurcevic}, \citenamefont {Muschik},
  \citenamefont {Silvi}, \citenamefont {Blatt}, \citenamefont {Roos},\ and\
  \citenamefont {Zoller}}]{Kokail2019}%
  \BibitemOpen
  \bibfield  {author} {\bibinfo {author} {\bibfnamefont {C.}~\bibnamefont
  {Kokail}}, \bibinfo {author} {\bibfnamefont {C.}~\bibnamefont {Maier}},
  \bibinfo {author} {\bibfnamefont {R.}~\bibnamefont {van Bijnen}}, \bibinfo
  {author} {\bibfnamefont {T.}~\bibnamefont {Brydges}}, \bibinfo {author}
  {\bibfnamefont {M.~K.}\ \bibnamefont {Joshi}}, \bibinfo {author}
  {\bibfnamefont {P.}~\bibnamefont {Jurcevic}}, \bibinfo {author}
  {\bibfnamefont {C.~A.}\ \bibnamefont {Muschik}}, \bibinfo {author}
  {\bibfnamefont {P.}~\bibnamefont {Silvi}}, \bibinfo {author} {\bibfnamefont
  {R.}~\bibnamefont {Blatt}}, \bibinfo {author} {\bibfnamefont {C.~F.}\
  \bibnamefont {Roos}}, \ and\ \bibinfo {author} {\bibfnamefont
  {P.}~\bibnamefont {Zoller}},\ }\href {\doibase 10.1038/s41586-019-1177-4}
  {\bibfield  {journal} {\bibinfo  {journal} {Nature}\ }\textbf {\bibinfo
  {volume} {569}},\ \bibinfo {pages} {355} (\bibinfo {year}
  {2019})}\BibitemShut {NoStop}%
\bibitem [{\citenamefont {Kandala}\ \emph {et~al.}(2017)\citenamefont
  {Kandala}, \citenamefont {Mezzacapo}, \citenamefont {Temme}, \citenamefont
  {Takita}, \citenamefont {Brink}, \citenamefont {Chow},\ and\ \citenamefont
  {Gambetta}}]{Kandala2017}%
  \BibitemOpen
  \bibfield  {author} {\bibinfo {author} {\bibfnamefont {A.}~\bibnamefont
  {Kandala}}, \bibinfo {author} {\bibfnamefont {A.}~\bibnamefont {Mezzacapo}},
  \bibinfo {author} {\bibfnamefont {K.}~\bibnamefont {Temme}}, \bibinfo
  {author} {\bibfnamefont {M.}~\bibnamefont {Takita}}, \bibinfo {author}
  {\bibfnamefont {M.}~\bibnamefont {Brink}}, \bibinfo {author} {\bibfnamefont
  {J.~M.}\ \bibnamefont {Chow}}, \ and\ \bibinfo {author} {\bibfnamefont
  {J.~M.}\ \bibnamefont {Gambetta}},\ }\href {\doibase 10.1038/nature23879}
  {\bibfield  {journal} {\bibinfo  {journal} {Nature}\ }\textbf {\bibinfo
  {volume} {549}},\ \bibinfo {pages} {242} (\bibinfo {year}
  {2017})}\BibitemShut {NoStop}%
\bibitem [{\citenamefont {Huggins}\ \emph {et~al.}(2019)\citenamefont
  {Huggins}, \citenamefont {Patil}, \citenamefont {Mitchell}, \citenamefont
  {{Birgitta Whaley}},\ and\ \citenamefont {{Miles
  Stoudenmire}}}]{Huggins2019}%
  \BibitemOpen
  \bibfield  {author} {\bibinfo {author} {\bibfnamefont {W.}~\bibnamefont
  {Huggins}}, \bibinfo {author} {\bibfnamefont {P.}~\bibnamefont {Patil}},
  \bibinfo {author} {\bibfnamefont {B.}~\bibnamefont {Mitchell}}, \bibinfo
  {author} {\bibfnamefont {K.}~\bibnamefont {{Birgitta Whaley}}}, \ and\
  \bibinfo {author} {\bibfnamefont {E.}~\bibnamefont {{Miles Stoudenmire}}},\
  }\href {\doibase 10.1088/2058-9565/aaea94} {\bibfield  {journal} {\bibinfo
  {journal} {Quantum Science and Technology}\ }\textbf {\bibinfo {volume}
  {4}},\ \bibinfo {pages} {024001} (\bibinfo {year} {2019})}\BibitemShut
  {NoStop}%
\bibitem [{\citenamefont {Liu}\ \emph {et~al.}(2019)\citenamefont {Liu},
  \citenamefont {Zhang}, \citenamefont {Wan},\ and\ \citenamefont
  {Wang}}]{Liu2019}%
  \BibitemOpen
  \bibfield  {author} {\bibinfo {author} {\bibfnamefont {J.-G.}\ \bibnamefont
  {Liu}}, \bibinfo {author} {\bibfnamefont {Y.-H.}\ \bibnamefont {Zhang}},
  \bibinfo {author} {\bibfnamefont {Y.}~\bibnamefont {Wan}}, \ and\ \bibinfo
  {author} {\bibfnamefont {L.}~\bibnamefont {Wang}},\ }\href {\doibase
  10.1103/PhysRevResearch.1.023025} {\bibfield  {journal} {\bibinfo  {journal}
  {Physical Review Research}\ }\textbf {\bibinfo {volume} {1}},\ \bibinfo
  {pages} {023025} (\bibinfo {year} {2019})}\BibitemShut {NoStop}%
\bibitem [{\citenamefont {Fujii}\ \emph {et~al.}(2020)\citenamefont {Fujii},
  \citenamefont {Mitarai}, \citenamefont {Mizukami},\ and\ \citenamefont
  {Nakagawa}}]{Fujii2020}%
  \BibitemOpen
  \bibfield  {author} {\bibinfo {author} {\bibfnamefont {K.}~\bibnamefont
  {Fujii}}, \bibinfo {author} {\bibfnamefont {K.}~\bibnamefont {Mitarai}},
  \bibinfo {author} {\bibfnamefont {W.}~\bibnamefont {Mizukami}}, \ and\
  \bibinfo {author} {\bibfnamefont {Y.~O.}\ \bibnamefont {Nakagawa}},\ }\href
  {http://arxiv.org/abs/2007.10917} {\  (\bibinfo {year} {2020})},\ \Eprint
  {http://arxiv.org/abs/2007.10917} {arXiv:2007.10917} \BibitemShut {NoStop}%
\bibitem [{\citenamefont {Koczor}\ and\ \citenamefont
  {Benjamin}(2021)}]{Koczor2020}%
  \BibitemOpen
  \bibfield  {author} {\bibinfo {author} {\bibfnamefont {B.}~\bibnamefont
  {Koczor}}\ and\ \bibinfo {author} {\bibfnamefont {S.~C.}\ \bibnamefont
  {Benjamin}},\ }\href@noop {} {} (\bibinfo {year} {2021}),\ \Eprint
  {http://arxiv.org/abs/2008.13774} {arXiv:2008.13774} \BibitemShut {NoStop}%
\bibitem [{\citenamefont {Ostaszewski}\ \emph {et~al.}(2021)\citenamefont
  {Ostaszewski}, \citenamefont {Grant},\ and\ \citenamefont
  {Benedetti}}]{Ostaszewski2021}%
  \BibitemOpen
  \bibfield  {author} {\bibinfo {author} {\bibfnamefont {M.}~\bibnamefont
  {Ostaszewski}}, \bibinfo {author} {\bibfnamefont {E.}~\bibnamefont {Grant}},
  \ and\ \bibinfo {author} {\bibfnamefont {M.}~\bibnamefont {Benedetti}},\
  }\href {\doibase 10.22331/q-2021-01-28-391} {\bibfield  {journal} {\bibinfo
  {journal} {Quantum}\ }\textbf {\bibinfo {volume} {5}},\ \bibinfo {pages}
  {391} (\bibinfo {year} {2021})}\BibitemShut {NoStop}%
\bibitem [{\citenamefont {Zeng}\ \emph {et~al.}(2021)\citenamefont {Zeng},
  \citenamefont {Wu}, \citenamefont {Cao}, \citenamefont {Zhang}, \citenamefont
  {Hou}, \citenamefont {Xu},\ and\ \citenamefont {Zeng}}]{Zeng2020}%
  \BibitemOpen
  \bibfield  {author} {\bibinfo {author} {\bibfnamefont {J.}~\bibnamefont
  {Zeng}}, \bibinfo {author} {\bibfnamefont {Z.}~\bibnamefont {Wu}}, \bibinfo
  {author} {\bibfnamefont {C.}~\bibnamefont {Cao}}, \bibinfo {author}
  {\bibfnamefont {C.}~\bibnamefont {Zhang}}, \bibinfo {author} {\bibfnamefont
  {S.-Y.}\ \bibnamefont {Hou}}, \bibinfo {author} {\bibfnamefont
  {P.}~\bibnamefont {Xu}}, \ and\ \bibinfo {author} {\bibfnamefont
  {B.}~\bibnamefont {Zeng}},\ }\href@noop {} {\bibfield  {journal} {\bibinfo
  {journal} {Quantum Engineering}\ }\textbf {\bibinfo {volume} {3}},\ \bibinfo
  {pages} {e77} (\bibinfo {year} {2021})}\BibitemShut {NoStop}%
\bibitem [{\citenamefont {Bethe}(1931)}]{Bethe1931}%
  \BibitemOpen
  \bibfield  {author} {\bibinfo {author} {\bibfnamefont {H.}~\bibnamefont
  {Bethe}},\ }\href {\doibase 10.1007/BF01341708} {\bibfield  {journal}
  {\bibinfo  {journal} {Zeitschrift f{\"{u}}r Physik}\ }\textbf {\bibinfo
  {volume} {71}},\ \bibinfo {pages} {205} (\bibinfo {year} {1931})}\BibitemShut
  {NoStop}%
\bibitem [{\citenamefont {Nepomechie}(2020)}]{Nepomechie2020}%
  \BibitemOpen
  \bibfield  {author} {\bibinfo {author} {\bibfnamefont {R.~I.}\ \bibnamefont
  {Nepomechie}},\ }\href {http://arxiv.org/abs/2010.01609} {\  (\bibinfo {year}
  {2020})},\ \Eprint {http://arxiv.org/abs/2010.01609} {arXiv:2010.01609}
  \BibitemShut {NoStop}%
\bibitem [{\citenamefont {{Van Dyke}}\ \emph {et~al.}(2021)\citenamefont {{Van
  Dyke}}, \citenamefont {Barron}, \citenamefont {Mayhall}, \citenamefont
  {Barnes},\ and\ \citenamefont {Economou}}]{VanDyke2021}%
  \BibitemOpen
  \bibfield  {author} {\bibinfo {author} {\bibfnamefont {J.~S.}\ \bibnamefont
  {{Van Dyke}}}, \bibinfo {author} {\bibfnamefont {G.~S.}\ \bibnamefont
  {Barron}}, \bibinfo {author} {\bibfnamefont {N.~J.}\ \bibnamefont {Mayhall}},
  \bibinfo {author} {\bibfnamefont {E.}~\bibnamefont {Barnes}}, \ and\ \bibinfo
  {author} {\bibfnamefont {S.~E.}\ \bibnamefont {Economou}},\ }\href
  {http://arxiv.org/abs/2103.13388} {\  (\bibinfo {year} {2021})},\ \Eprint
  {http://arxiv.org/abs/2103.13388} {arXiv:2103.13388} \BibitemShut {NoStop}%
\bibitem [{\citenamefont {Lyu}\ \emph {et~al.}(2020)\citenamefont {Lyu},
  \citenamefont {Montenegro},\ and\ \citenamefont {Bayat}}]{Lyu2020}%
  \BibitemOpen
  \bibfield  {author} {\bibinfo {author} {\bibfnamefont {C.}~\bibnamefont
  {Lyu}}, \bibinfo {author} {\bibfnamefont {V.}~\bibnamefont {Montenegro}}, \
  and\ \bibinfo {author} {\bibfnamefont {A.}~\bibnamefont {Bayat}},\ }\href
  {\doibase 10.22331/q-2020-09-16-324} {\bibfield  {journal} {\bibinfo
  {journal} {Quantum}\ }\textbf {\bibinfo {volume} {4}},\ \bibinfo {pages}
  {324} (\bibinfo {year} {2020})}\BibitemShut {NoStop}%
\bibitem [{\citenamefont {Seki}\ \emph {et~al.}(2020)\citenamefont {Seki},
  \citenamefont {Shirakawa},\ and\ \citenamefont {Yunoki}}]{Seki2020}%
  \BibitemOpen
  \bibfield  {author} {\bibinfo {author} {\bibfnamefont {K.}~\bibnamefont
  {Seki}}, \bibinfo {author} {\bibfnamefont {T.}~\bibnamefont {Shirakawa}}, \
  and\ \bibinfo {author} {\bibfnamefont {S.}~\bibnamefont {Yunoki}},\ }\href
  {\doibase 10.1103/PhysRevA.101.052340} {\bibfield  {journal} {\bibinfo
  {journal} {Physical Review A}\ }\textbf {\bibinfo {volume} {101}},\ \bibinfo
  {pages} {052340} (\bibinfo {year} {2020})}\BibitemShut {NoStop}%
\bibitem [{\citenamefont {Slattery}\ \emph {et~al.}(2021)\citenamefont
  {Slattery}, \citenamefont {Villalonga},\ and\ \citenamefont
  {Clark}}]{Slattery2021}%
  \BibitemOpen
  \bibfield  {author} {\bibinfo {author} {\bibfnamefont {L.}~\bibnamefont
  {Slattery}}, \bibinfo {author} {\bibfnamefont {B.}~\bibnamefont
  {Villalonga}}, \ and\ \bibinfo {author} {\bibfnamefont {B.~K.}\ \bibnamefont
  {Clark}},\ }\href {http://arxiv.org/abs/2102.08403} {\  (\bibinfo {year}
  {2021})},\ \Eprint {http://arxiv.org/abs/2102.08403} {arXiv:2102.08403}
  \BibitemShut {NoStop}%
\bibitem [{\citenamefont {Jin}\ \emph {et~al.}(2020)\citenamefont {Jin},
  \citenamefont {Wu}, \citenamefont {Zhou}, \citenamefont {Li}, \citenamefont
  {Li}, \citenamefont {Li},\ and\ \citenamefont {Wang}}]{Jin2020}%
  \BibitemOpen
  \bibfield  {author} {\bibinfo {author} {\bibfnamefont {S.}~\bibnamefont
  {Jin}}, \bibinfo {author} {\bibfnamefont {S.}~\bibnamefont {Wu}}, \bibinfo
  {author} {\bibfnamefont {G.}~\bibnamefont {Zhou}}, \bibinfo {author}
  {\bibfnamefont {Y.}~\bibnamefont {Li}}, \bibinfo {author} {\bibfnamefont
  {L.}~\bibnamefont {Li}}, \bibinfo {author} {\bibfnamefont {B.}~\bibnamefont
  {Li}}, \ and\ \bibinfo {author} {\bibfnamefont {X.}~\bibnamefont {Wang}},\
  }\href {\doibase https://doi.org/10.1002/que2.49} {\bibfield  {journal}
  {\bibinfo  {journal} {Quantum Engineering}\ }\textbf {\bibinfo {volume}
  {2}},\ \bibinfo {pages} {e49} (\bibinfo {year} {2020})}\BibitemShut {NoStop}%
\bibitem [{\citenamefont {Bespalova}\ and\ \citenamefont
  {Kyriienko}(2021)}]{Bespalova2020}%
  \BibitemOpen
  \bibfield  {author} {\bibinfo {author} {\bibfnamefont {T.~A.}\ \bibnamefont
  {Bespalova}}\ and\ \bibinfo {author} {\bibfnamefont {O.}~\bibnamefont
  {Kyriienko}},\ }\href {\doibase 10.1103/PRXQuantum.2.030318} {\bibfield
  {journal} {\bibinfo  {journal} {PRX Quantum}\ }\textbf {\bibinfo {volume}
  {2}},\ \bibinfo {pages} {030318} (\bibinfo {year} {2021})}\BibitemShut
  {NoStop}%
\bibitem [{\citenamefont {Grimsley}\ \emph {et~al.}(2020)\citenamefont
  {Grimsley}, \citenamefont {Claudino}, \citenamefont {Economou}, \citenamefont
  {Barnes},\ and\ \citenamefont {Mayhall}}]{Grimsley2020}%
  \BibitemOpen
  \bibfield  {author} {\bibinfo {author} {\bibfnamefont {H.~R.}\ \bibnamefont
  {Grimsley}}, \bibinfo {author} {\bibfnamefont {D.}~\bibnamefont {Claudino}},
  \bibinfo {author} {\bibfnamefont {S.~E.}\ \bibnamefont {Economou}}, \bibinfo
  {author} {\bibfnamefont {E.}~\bibnamefont {Barnes}}, \ and\ \bibinfo {author}
  {\bibfnamefont {N.~J.}\ \bibnamefont {Mayhall}},\ }\href {\doibase
  10.1021/acs.jctc.9b01083} {\bibfield  {journal} {\bibinfo  {journal} {Journal
  of Chemical Theory and Computation}\ }\textbf {\bibinfo {volume} {16}},\
  \bibinfo {pages} {1} (\bibinfo {year} {2020})}\BibitemShut {NoStop}%
\bibitem [{\citenamefont {Tranter}\ \emph {et~al.}(2019)\citenamefont
  {Tranter}, \citenamefont {Love}, \citenamefont {Mintert}, \citenamefont
  {Wiebe},\ and\ \citenamefont {Coveney}}]{Tranter2019}%
  \BibitemOpen
  \bibfield  {author} {\bibinfo {author} {\bibfnamefont {A.}~\bibnamefont
  {Tranter}}, \bibinfo {author} {\bibfnamefont {P.~J.}\ \bibnamefont {Love}},
  \bibinfo {author} {\bibfnamefont {F.}~\bibnamefont {Mintert}}, \bibinfo
  {author} {\bibfnamefont {N.}~\bibnamefont {Wiebe}}, \ and\ \bibinfo {author}
  {\bibfnamefont {P.~V.}\ \bibnamefont {Coveney}},\ }\href@noop {} {\bibfield
  {journal} {\bibinfo  {journal} {Entropy}\ }\textbf {\bibinfo {volume} {21}},\
  \bibinfo {pages} {1218} (\bibinfo {year} {2019})}\BibitemShut {NoStop}%
\bibitem [{\citenamefont {{De Raedt}}(1987)}]{DeRaedt1987}%
  \BibitemOpen
  \bibfield  {author} {\bibinfo {author} {\bibfnamefont {H.}~\bibnamefont {{De
  Raedt}}},\ }\href {\doibase 10.1016/0167-7977(87)90002-5} {\bibfield
  {journal} {\bibinfo  {journal} {Computer Physics Reports}\ }\textbf {\bibinfo
  {volume} {7}},\ \bibinfo {pages} {1} (\bibinfo {year} {1987})}\BibitemShut
  {NoStop}%
\bibitem [{\citenamefont {Xia}\ and\ \citenamefont {Kais}(2020)}]{Xia2020}%
  \BibitemOpen
  \bibfield  {author} {\bibinfo {author} {\bibfnamefont {R.}~\bibnamefont
  {Xia}}\ and\ \bibinfo {author} {\bibfnamefont {S.}~\bibnamefont {Kais}},\
  }\href {\doibase 10.1088/2058-9565/abbc74} {\bibfield  {journal} {\bibinfo
  {journal} {Quantum Science and Technology}\ }\textbf {\bibinfo {volume}
  {6}},\ \bibinfo {pages} {015001} (\bibinfo {year} {2020})}\BibitemShut
  {NoStop}%
\bibitem [{\citenamefont {Romero}\ \emph {et~al.}(2019)\citenamefont {Romero},
  \citenamefont {Babbush}, \citenamefont {McClean}, \citenamefont {Hempel},
  \citenamefont {Love},\ and\ \citenamefont {Aspuru-Guzik}}]{Romero2019}%
  \BibitemOpen
  \bibfield  {author} {\bibinfo {author} {\bibfnamefont {J.}~\bibnamefont
  {Romero}}, \bibinfo {author} {\bibfnamefont {R.}~\bibnamefont {Babbush}},
  \bibinfo {author} {\bibfnamefont {J.~R.}\ \bibnamefont {McClean}}, \bibinfo
  {author} {\bibfnamefont {C.}~\bibnamefont {Hempel}}, \bibinfo {author}
  {\bibfnamefont {P.~J.}\ \bibnamefont {Love}}, \ and\ \bibinfo {author}
  {\bibfnamefont {A.}~\bibnamefont {Aspuru-Guzik}},\ }\href {\doibase
  10.1088/2058-9565/aad3e4} {\bibfield  {journal} {\bibinfo  {journal} {Quantum
  Science and Technology}\ }\textbf {\bibinfo {volume} {4}},\ \bibinfo {pages}
  {014008} (\bibinfo {year} {2019})}\BibitemShut {NoStop}%
\bibitem [{\citenamefont {Barkoutsos}\ \emph {et~al.}(2018)\citenamefont
  {Barkoutsos}, \citenamefont {Gonthier}, \citenamefont {Sokolov},
  \citenamefont {Moll}, \citenamefont {Salis}, \citenamefont {Fuhrer},
  \citenamefont {Ganzhorn}, \citenamefont {Egger}, \citenamefont {Troyer},
  \citenamefont {Mezzacapo}, \citenamefont {Filipp},\ and\ \citenamefont
  {Tavernelli}}]{Barkoutsos2018}%
  \BibitemOpen
  \bibfield  {author} {\bibinfo {author} {\bibfnamefont {P.~K.}\ \bibnamefont
  {Barkoutsos}}, \bibinfo {author} {\bibfnamefont {J.~F.}\ \bibnamefont
  {Gonthier}}, \bibinfo {author} {\bibfnamefont {I.}~\bibnamefont {Sokolov}},
  \bibinfo {author} {\bibfnamefont {N.}~\bibnamefont {Moll}}, \bibinfo {author}
  {\bibfnamefont {G.}~\bibnamefont {Salis}}, \bibinfo {author} {\bibfnamefont
  {A.}~\bibnamefont {Fuhrer}}, \bibinfo {author} {\bibfnamefont
  {M.}~\bibnamefont {Ganzhorn}}, \bibinfo {author} {\bibfnamefont {D.~J.}\
  \bibnamefont {Egger}}, \bibinfo {author} {\bibfnamefont {M.}~\bibnamefont
  {Troyer}}, \bibinfo {author} {\bibfnamefont {A.}~\bibnamefont {Mezzacapo}},
  \bibinfo {author} {\bibfnamefont {S.}~\bibnamefont {Filipp}}, \ and\ \bibinfo
  {author} {\bibfnamefont {I.}~\bibnamefont {Tavernelli}},\ }\href {\doibase
  10.1103/PhysRevA.98.022322} {\bibfield  {journal} {\bibinfo  {journal}
  {Physical Review A}\ }\textbf {\bibinfo {volume} {98}},\ \bibinfo {pages}
  {022322} (\bibinfo {year} {2018})}\BibitemShut {NoStop}%
\bibitem [{\citenamefont {Shen}\ \emph {et~al.}(2017)\citenamefont {Shen},
  \citenamefont {Zhang}, \citenamefont {Zhang}, \citenamefont {Zhang},
  \citenamefont {Yung},\ and\ \citenamefont {Kim}}]{Shen2017}%
  \BibitemOpen
  \bibfield  {author} {\bibinfo {author} {\bibfnamefont {Y.}~\bibnamefont
  {Shen}}, \bibinfo {author} {\bibfnamefont {X.}~\bibnamefont {Zhang}},
  \bibinfo {author} {\bibfnamefont {S.}~\bibnamefont {Zhang}}, \bibinfo
  {author} {\bibfnamefont {J.~N.}\ \bibnamefont {Zhang}}, \bibinfo {author}
  {\bibfnamefont {M.~H.}\ \bibnamefont {Yung}}, \ and\ \bibinfo {author}
  {\bibfnamefont {K.}~\bibnamefont {Kim}},\ }\href {\doibase
  10.1103/PhysRevA.95.020501} {\bibfield  {journal} {\bibinfo  {journal}
  {Physical Review A}\ }\textbf {\bibinfo {volume} {95}},\ \bibinfo {pages}
  {020501} (\bibinfo {year} {2017})}\BibitemShut {NoStop}%
\bibitem [{\citenamefont {Wecker}\ \emph {et~al.}(2015)\citenamefont {Wecker},
  \citenamefont {Hastings},\ and\ \citenamefont {Troyer}}]{Wecker2015}%
  \BibitemOpen
  \bibfield  {author} {\bibinfo {author} {\bibfnamefont {D.}~\bibnamefont
  {Wecker}}, \bibinfo {author} {\bibfnamefont {M.~B.}\ \bibnamefont
  {Hastings}}, \ and\ \bibinfo {author} {\bibfnamefont {M.}~\bibnamefont
  {Troyer}},\ }\href {\doibase 10.1103/PhysRevA.92.042303} {\bibfield
  {journal} {\bibinfo  {journal} {Phys. Rev. A}\ }\textbf {\bibinfo {volume}
  {92}},\ \bibinfo {pages} {042303} (\bibinfo {year} {2015})}\BibitemShut
  {NoStop}%
\bibitem [{\citenamefont {Reiner}\ \emph {et~al.}(2019)\citenamefont {Reiner},
  \citenamefont {Wilhelm-Mauch}, \citenamefont {Schön},\ and\ \citenamefont
  {Marthaler}}]{Reiner2019}%
  \BibitemOpen
  \bibfield  {author} {\bibinfo {author} {\bibfnamefont {J.-M.}\ \bibnamefont
  {Reiner}}, \bibinfo {author} {\bibfnamefont {F.}~\bibnamefont
  {Wilhelm-Mauch}}, \bibinfo {author} {\bibfnamefont {G.}~\bibnamefont
  {Schön}}, \ and\ \bibinfo {author} {\bibfnamefont {M.}~\bibnamefont
  {Marthaler}},\ }\href {\doibase 10.1088/2058-9565/ab1e85} {\bibfield
  {journal} {\bibinfo  {journal} {Quantum Science and Technology}\ }\textbf
  {\bibinfo {volume} {4}},\ \bibinfo {pages} {035005} (\bibinfo {year}
  {2019})}\BibitemShut {NoStop}%
\bibitem [{\citenamefont {Choquette}\ \emph {et~al.}(2021)\citenamefont
  {Choquette}, \citenamefont {Di~Paolo}, \citenamefont {Barkoutsos},
  \citenamefont {S\'en\'echal}, \citenamefont {Tavernelli},\ and\ \citenamefont
  {Blais}}]{okokok}%
  \BibitemOpen
  \bibfield  {author} {\bibinfo {author} {\bibfnamefont {A.}~\bibnamefont
  {Choquette}}, \bibinfo {author} {\bibfnamefont {A.}~\bibnamefont {Di~Paolo}},
  \bibinfo {author} {\bibfnamefont {P.~K.}\ \bibnamefont {Barkoutsos}},
  \bibinfo {author} {\bibfnamefont {D.}~\bibnamefont {S\'en\'echal}}, \bibinfo
  {author} {\bibfnamefont {I.}~\bibnamefont {Tavernelli}}, \ and\ \bibinfo
  {author} {\bibfnamefont {A.}~\bibnamefont {Blais}},\ }\href {\doibase
  10.1103/PhysRevResearch.3.023092} {\bibfield  {journal} {\bibinfo  {journal}
  {Phys. Rev. Research}\ }\textbf {\bibinfo {volume} {3}},\ \bibinfo {pages}
  {023092} (\bibinfo {year} {2021})}\BibitemShut {NoStop}%
\bibitem [{\citenamefont {Wiersema}\ \emph {et~al.}(2020)\citenamefont
  {Wiersema}, \citenamefont {Zhou}, \citenamefont {de~Sereville}, \citenamefont
  {Carrasquilla}, \citenamefont {Kim},\ and\ \citenamefont
  {Yuen}}]{Wiersema2020}%
  \BibitemOpen
  \bibfield  {author} {\bibinfo {author} {\bibfnamefont {R.}~\bibnamefont
  {Wiersema}}, \bibinfo {author} {\bibfnamefont {C.}~\bibnamefont {Zhou}},
  \bibinfo {author} {\bibfnamefont {Y.}~\bibnamefont {de~Sereville}}, \bibinfo
  {author} {\bibfnamefont {J.~F.}\ \bibnamefont {Carrasquilla}}, \bibinfo
  {author} {\bibfnamefont {Y.~B.}\ \bibnamefont {Kim}}, \ and\ \bibinfo
  {author} {\bibfnamefont {H.}~\bibnamefont {Yuen}},\ }\href {\doibase
  10.1103/PRXQuantum.1.020319} {\bibfield  {journal} {\bibinfo  {journal} {PRX
  Quantum}\ }\textbf {\bibinfo {volume} {1}},\ \bibinfo {pages} {020319}
  (\bibinfo {year} {2020})}\BibitemShut {NoStop}%
\bibitem [{\citenamefont {{De Raedt}}\ \emph {et~al.}(2007)\citenamefont {{De
  Raedt}}, \citenamefont {Michielsen}, \citenamefont {{De Raedt}},
  \citenamefont {Trieu}, \citenamefont {Arnold}, \citenamefont {Richter},
  \citenamefont {{Th. Lippert}}, \citenamefont {Watanabe},\ and\ \citenamefont
  {Ito}}]{DeRaedt2007}%
  \BibitemOpen
  \bibfield  {author} {\bibinfo {author} {\bibfnamefont {K.}~\bibnamefont {{De
  Raedt}}}, \bibinfo {author} {\bibfnamefont {K.}~\bibnamefont {Michielsen}},
  \bibinfo {author} {\bibfnamefont {H.}~\bibnamefont {{De Raedt}}}, \bibinfo
  {author} {\bibfnamefont {B.}~\bibnamefont {Trieu}}, \bibinfo {author}
  {\bibfnamefont {G.}~\bibnamefont {Arnold}}, \bibinfo {author} {\bibfnamefont
  {M.}~\bibnamefont {Richter}}, \bibinfo {author} {\bibnamefont {{Th.
  Lippert}}}, \bibinfo {author} {\bibfnamefont {H.}~\bibnamefont {Watanabe}}, \
  and\ \bibinfo {author} {\bibfnamefont {N.}~\bibnamefont {Ito}},\ }\href
  {\doibase 10.1016/j.cpc.2006.08.007} {\bibfield  {journal} {\bibinfo
  {journal} {Computer Physics Communications}\ }\textbf {\bibinfo {volume}
  {176}},\ \bibinfo {pages} {121} (\bibinfo {year} {2007})}\BibitemShut
  {NoStop}%
\bibitem [{\citenamefont {{De Raedt}}\ \emph {et~al.}(2019)\citenamefont {{De
  Raedt}}, \citenamefont {Jin}, \citenamefont {Willsch}, \citenamefont
  {Willsch}, \citenamefont {Yoshioka}, \citenamefont {Ito}, \citenamefont
  {Yuan},\ and\ \citenamefont {Michielsen}}]{DeRaedt2019}%
  \BibitemOpen
  \bibfield  {author} {\bibinfo {author} {\bibfnamefont {H.}~\bibnamefont {{De
  Raedt}}}, \bibinfo {author} {\bibfnamefont {F.}~\bibnamefont {Jin}}, \bibinfo
  {author} {\bibfnamefont {D.}~\bibnamefont {Willsch}}, \bibinfo {author}
  {\bibfnamefont {M.}~\bibnamefont {Willsch}}, \bibinfo {author} {\bibfnamefont
  {N.}~\bibnamefont {Yoshioka}}, \bibinfo {author} {\bibfnamefont
  {N.}~\bibnamefont {Ito}}, \bibinfo {author} {\bibfnamefont {S.}~\bibnamefont
  {Yuan}}, \ and\ \bibinfo {author} {\bibfnamefont {K.}~\bibnamefont
  {Michielsen}},\ }\href {\doibase https://doi.org/10.1016/j.cpc.2018.11.005}
  {\bibfield  {journal} {\bibinfo  {journal} {Computer Physics Communications}\
  }\textbf {\bibinfo {volume} {237}},\ \bibinfo {pages} {47} (\bibinfo {year}
  {2019})}\BibitemShut {NoStop}%
\bibitem [{\citenamefont {{G. Aleksandrowicz et al.}}(2019)}]{Qiskit}%
  \BibitemOpen
  \bibfield  {author} {\bibinfo {author} {\bibnamefont {{G. Aleksandrowicz et
  al.}}},\ }\href {\doibase 10.5281/zenodo.2562110} {\enquote {\bibinfo {title}
  {Qiskit: An open-source framework for quantum computing},}\ } (\bibinfo
  {year} {2019})\BibitemShut {NoStop}%
\bibitem [{\citenamefont {Bonet-Monroig}\ \emph {et~al.}(2020)\citenamefont
  {Bonet-Monroig}, \citenamefont {Babbush},\ and\ \citenamefont
  {O'Brien}}]{Bonet-Monroig2020}%
  \BibitemOpen
  \bibfield  {author} {\bibinfo {author} {\bibfnamefont {X.}~\bibnamefont
  {Bonet-Monroig}}, \bibinfo {author} {\bibfnamefont {R.}~\bibnamefont
  {Babbush}}, \ and\ \bibinfo {author} {\bibfnamefont {T.~E.}\ \bibnamefont
  {O'Brien}},\ }\href {\doibase 10.1103/PhysRevX.10.031064} {\bibfield
  {journal} {\bibinfo  {journal} {Physical Review X}\ }\textbf {\bibinfo
  {volume} {10}},\ \bibinfo {pages} {031064} (\bibinfo {year}
  {2020})}\BibitemShut {NoStop}%
\bibitem [{\citenamefont {Verteletskyi}\ \emph {et~al.}(2020)\citenamefont
  {Verteletskyi}, \citenamefont {Yen},\ and\ \citenamefont
  {Izmaylov}}]{Verteletskyi2020}%
  \BibitemOpen
  \bibfield  {author} {\bibinfo {author} {\bibfnamefont {V.}~\bibnamefont
  {Verteletskyi}}, \bibinfo {author} {\bibfnamefont {T.~C.}\ \bibnamefont
  {Yen}}, \ and\ \bibinfo {author} {\bibfnamefont {A.~F.}\ \bibnamefont
  {Izmaylov}},\ }\href {\doibase 10.1063/1.5141458} {\bibfield  {journal}
  {\bibinfo  {journal} {Journal of Chemical Physics}\ }\textbf {\bibinfo
  {volume} {152}},\ \bibinfo {pages} {224109} (\bibinfo {year}
  {2020})}\BibitemShut {NoStop}%
\bibitem [{\citenamefont {Hadfield}\ \emph {et~al.}(2020)\citenamefont
  {Hadfield}, \citenamefont {Bravyi}, \citenamefont {Raymond},\ and\
  \citenamefont {Mezzacapo}}]{Hadfield2020}%
  \BibitemOpen
  \bibfield  {author} {\bibinfo {author} {\bibfnamefont {C.}~\bibnamefont
  {Hadfield}}, \bibinfo {author} {\bibfnamefont {S.}~\bibnamefont {Bravyi}},
  \bibinfo {author} {\bibfnamefont {R.}~\bibnamefont {Raymond}}, \ and\
  \bibinfo {author} {\bibfnamefont {A.}~\bibnamefont {Mezzacapo}},\ }\href@noop
  {} {\  (\bibinfo {year} {2020})},\ \Eprint {http://arxiv.org/abs/2006.15788}
  {arXiv:2006.15788} \BibitemShut {NoStop}%
\bibitem [{\citenamefont {Gokhale}\ \emph {et~al.}(2020)\citenamefont
  {Gokhale}, \citenamefont {Angiuli}, \citenamefont {Ding}, \citenamefont
  {Gui}, \citenamefont {Tomesh}, \citenamefont {Suchara}, \citenamefont
  {Martonosi},\ and\ \citenamefont {Chong}}]{Gokhale2019}%
  \BibitemOpen
  \bibfield  {author} {\bibinfo {author} {\bibfnamefont {P.}~\bibnamefont
  {Gokhale}}, \bibinfo {author} {\bibfnamefont {O.}~\bibnamefont {Angiuli}},
  \bibinfo {author} {\bibfnamefont {Y.}~\bibnamefont {Ding}}, \bibinfo {author}
  {\bibfnamefont {K.}~\bibnamefont {Gui}}, \bibinfo {author} {\bibfnamefont
  {T.}~\bibnamefont {Tomesh}}, \bibinfo {author} {\bibfnamefont
  {M.}~\bibnamefont {Suchara}}, \bibinfo {author} {\bibfnamefont
  {M.}~\bibnamefont {Martonosi}}, \ and\ \bibinfo {author} {\bibfnamefont
  {F.~T.}\ \bibnamefont {Chong}},\ }\href {\doibase 10.1109/TQE.2020.3035814}
  {\bibfield  {journal} {\bibinfo  {journal} {IEEE Transactions on Quantum
  Engineering}\ }\textbf {\bibinfo {volume} {1}},\ \bibinfo {pages} {1}
  (\bibinfo {year} {2020})}\BibitemShut {NoStop}%
\bibitem [{\citenamefont {Huggins}\ \emph {et~al.}(2021)\citenamefont
  {Huggins}, \citenamefont {McClean}, \citenamefont {Rubin}, \citenamefont
  {Jiang}, \citenamefont {Wiebe}, \citenamefont {Whaley},\ and\ \citenamefont
  {Babbush}}]{Huggins2021}%
  \BibitemOpen
  \bibfield  {author} {\bibinfo {author} {\bibfnamefont {W.~J.}\ \bibnamefont
  {Huggins}}, \bibinfo {author} {\bibfnamefont {J.~R.}\ \bibnamefont
  {McClean}}, \bibinfo {author} {\bibfnamefont {N.~C.}\ \bibnamefont {Rubin}},
  \bibinfo {author} {\bibfnamefont {Z.}~\bibnamefont {Jiang}}, \bibinfo
  {author} {\bibfnamefont {N.}~\bibnamefont {Wiebe}}, \bibinfo {author}
  {\bibfnamefont {K.~B.}\ \bibnamefont {Whaley}}, \ and\ \bibinfo {author}
  {\bibfnamefont {R.}~\bibnamefont {Babbush}},\ }\href
  {https://doi.org/10.1038/s41534-020-00341-7} {\bibfield  {journal} {\bibinfo
  {journal} {npj Quantum Information}\ }\textbf {\bibinfo {volume} {7}},\
  \bibinfo {pages} {1} (\bibinfo {year} {2021})}\BibitemShut {NoStop}%
\bibitem [{\citenamefont {Kraft}(1988)}]{Kraft}%
  \BibitemOpen
  \bibfield  {author} {\bibinfo {author} {\bibfnamefont {D.}~\bibnamefont
  {Kraft}},\ }\href@noop {} {\emph {\bibinfo {title} {A software package for
  sequential quadratic programming}}}\ (\bibinfo {year} {1988})\BibitemShut
  {NoStop}%
\bibitem [{\citenamefont {{E. Jones, T. Oliphant, P. Peterson, et al. SciPy:
  Open source scientific tools for Python. http://www.scipy.org}}()}]{scipy}%
  \BibitemOpen
  \bibfield  {author} {\bibinfo {author} {\bibnamefont {{E. Jones, T. Oliphant,
  P. Peterson, et al. SciPy: Open source scientific tools for Python.
  http://www.scipy.org}}},\ }\href@noop {} {}\BibitemShut {NoStop}%
\bibitem [{\citenamefont {Powell}(1994)}]{Powell1994}%
  \BibitemOpen
  \bibfield  {author} {\bibinfo {author} {\bibfnamefont {M.~J.~D.}\
  \bibnamefont {Powell}},\ }in\ \href {\doibase 10.1007/978-94-015-8330-5_4}
  {\emph {\bibinfo {booktitle} {Advances in Optimization and Numerical
  Analysis}}}\ (\bibinfo  {publisher} {Springer Netherlands},\ \bibinfo {year}
  {1994})\ pp.\ \bibinfo {pages} {51--67}\BibitemShut {NoStop}%
\bibitem [{\citenamefont {Powell}(2007)}]{Powell2007}%
  \BibitemOpen
  \bibfield  {author} {\bibinfo {author} {\bibfnamefont {M.~J.~D.}\
  \bibnamefont {Powell}},\ }\href
  {http://www.damtp.cam.ac.uk/user/na/NA_papers/NA2007_03.pdf} {\bibfield
  {journal} {\bibinfo  {journal} {Mathematics Today}\ }\textbf {\bibinfo
  {volume} {43}},\ \bibinfo {pages} {170} (\bibinfo {year} {2007})}\BibitemShut
  {NoStop}%
\bibitem [{\citenamefont {{J\"{u}lich Supercomputing Centre. {JUWELS: Modular
  Tier-0/1 Supercomputer at the J\"{u}lich Supercomputing
  Centre}}}(2019)}]{JUWELS}%
  \BibitemOpen
  \bibfield  {author} {\bibinfo {author} {\bibnamefont {{J\"{u}lich
  Supercomputing Centre. {JUWELS: Modular Tier-0/1 Supercomputer at the
  J\"{u}lich Supercomputing Centre}}}},\ }\href
  {http://dx.doi.org/10.17815/jlsrf-5-171} {\bibfield  {journal} {\bibinfo
  {journal} {Journal of large-scale research facilities}\ }\textbf {\bibinfo
  {volume} {5}},\ \bibinfo {pages} {A135} (\bibinfo {year} {2019})}\BibitemShut
  {NoStop}%
\bibitem [{\citenamefont {McClean}\ \emph {et~al.}(2018)\citenamefont
  {McClean}, \citenamefont {Boixo}, \citenamefont {Smelyanskiy}, \citenamefont
  {Babbush},\ and\ \citenamefont {Neven}}]{McClean2018}%
  \BibitemOpen
  \bibfield  {author} {\bibinfo {author} {\bibfnamefont {J.~R.}\ \bibnamefont
  {McClean}}, \bibinfo {author} {\bibfnamefont {S.}~\bibnamefont {Boixo}},
  \bibinfo {author} {\bibfnamefont {V.~N.}\ \bibnamefont {Smelyanskiy}},
  \bibinfo {author} {\bibfnamefont {R.}~\bibnamefont {Babbush}}, \ and\
  \bibinfo {author} {\bibfnamefont {H.}~\bibnamefont {Neven}},\ }\href
  {\doibase 10.1038/s41467-018-07090-4} {\bibfield  {journal} {\bibinfo
  {journal} {Nature Communications}\ }\textbf {\bibinfo {volume} {9}},\
  \bibinfo {pages} {1} (\bibinfo {year} {2018})}\BibitemShut {NoStop}%
\bibitem [{\citenamefont {Campos}\ \emph {et~al.}(2021)\citenamefont {Campos},
  \citenamefont {Nasrallah},\ and\ \citenamefont {Biamonte}}]{Campos2021}%
  \BibitemOpen
  \bibfield  {author} {\bibinfo {author} {\bibfnamefont {E.}~\bibnamefont
  {Campos}}, \bibinfo {author} {\bibfnamefont {A.}~\bibnamefont {Nasrallah}}, \
  and\ \bibinfo {author} {\bibfnamefont {J.}~\bibnamefont {Biamonte}},\ }\href
  {\doibase 10.1103/PhysRevA.103.032607} {\bibfield  {journal} {\bibinfo
  {journal} {Physical Review A}\ }\textbf {\bibinfo {volume} {103}},\ \bibinfo
  {pages} {032607} (\bibinfo {year} {2021})}\BibitemShut {NoStop}%
\bibitem [{\citenamefont {Holmes}\ \emph {et~al.}(2021)\citenamefont {Holmes},
  \citenamefont {Sharma}, \citenamefont {Cerezo},\ and\ \citenamefont
  {Coles}}]{Holmes2021}%
  \BibitemOpen
  \bibfield  {author} {\bibinfo {author} {\bibfnamefont {Z.}~\bibnamefont
  {Holmes}}, \bibinfo {author} {\bibfnamefont {K.}~\bibnamefont {Sharma}},
  \bibinfo {author} {\bibfnamefont {M.}~\bibnamefont {Cerezo}}, \ and\ \bibinfo
  {author} {\bibfnamefont {P.~J.}\ \bibnamefont {Coles}},\ }\href
  {http://arxiv.org/abs/2101.02138} {\  (\bibinfo {year} {2021})},\ \Eprint
  {http://arxiv.org/abs/2101.02138} {arXiv:2101.02138} \BibitemShut {NoStop}%
\bibitem [{\citenamefont {Cerezo}\ and\ \citenamefont
  {Coles}(2021)}]{Cerezo2020}%
  \BibitemOpen
  \bibfield  {author} {\bibinfo {author} {\bibfnamefont {M.}~\bibnamefont
  {Cerezo}}\ and\ \bibinfo {author} {\bibfnamefont {P.~J.}\ \bibnamefont
  {Coles}},\ }\href {\doibase 10.1088/2058-9565/abf51a} {\bibfield  {journal}
  {\bibinfo  {journal} {Quantum Science and Technology}\ }\textbf {\bibinfo
  {volume} {6}},\ \bibinfo {pages} {035006} (\bibinfo {year}
  {2021})}\BibitemShut {NoStop}%
\bibitem [{\citenamefont {Parkinson}\ and\ \citenamefont
  {Farnell}(2010)}]{Parkinson2010}%
  \BibitemOpen
  \bibfield  {author} {\bibinfo {author} {\bibfnamefont {J.~B.}\ \bibnamefont
  {Parkinson}}\ and\ \bibinfo {author} {\bibfnamefont {D.~J.}\ \bibnamefont
  {Farnell}},\ }\href {\doibase 10.1007/978-3-642-13290-2_9} {\bibfield
  {journal} {\bibinfo  {journal} {Lecture Notes in Physics}\ }\textbf {\bibinfo
  {volume} {816}},\ \bibinfo {pages} {99} (\bibinfo {year} {2010})}\BibitemShut
  {NoStop}%
\end{thebibliography}%

\end{document}